\documentclass[aps,reprint,groupedaddress,nobibnotes]{revtex4-1}
\usepackage[utf8]{inputenc}
\usepackage{graphicx}
\graphicspath{ {../FIGURES_PAPER/} }
\usepackage{natbib}
\usepackage{amsmath,amssymb,amsfonts,amssymb}
\usepackage{subfig}
\usepackage{setspace}
\usepackage{float}
\usepackage{microtype}
\usepackage[normalem]{ulem}
\usepackage{ragged2e}
\usepackage{xcolor}

\DeclareCaptionJustification{justified}{\justifying}
\usepackage{hyperref}
\hypersetup{
	hyperindex,
	breaklinks,
	colorlinks=true,
	linkcolor=blue,
	citecolor=magenta,
	bookmarks=true,
	bookmarksopen=true,
	bookmarksopenlevel=2,
	pdfstartpage={1},
	pdfstartview={FitH},
	pdfview={FitH 0},
	pdfauthor={B. F. Farrell and P. J. Ioannou},
	pdftitle={Statistical state dynamics based theory for the formation and equilibration of Saturn's north polar jet}}

\usepackage{ifthen}

\def\U{\bm{\mathsf{U}}}

\def\C{\bm{\mathsf{C}}}
\def\I{\bm{\mathsf{I}}}

\def\R{{\bf R}}

\def\A{\bm{\mathsf{A}}}
\def\F{\bm{\mathsf{F}}}

\def\Q{\bm{\mathsf{Q}}}

\def\U{\bm{\mathsf{U}}}

\def\C{\bm{\mathsf{C}}}
\def\Q{\bm{\mathsf{Q}}}
\def\I{\bm{\mathsf{I}}}

\def\R{{\bf R}}

\def\A{\bm{\mathsf{A}}}
\def\F{\bm{\mathsf{F}}}

\def\Q{\bm{\mathsf{Q}}}

\newcommand\redsout{\bgroup\markoverwith{\textcolor{red}{\rule[0.5ex]{2pt}{0.4pt}}}\ULon}

\newcommand{\be}{\begin{equation}}
\newcommand{\ee}{\end{equation}}
\newcommand{\bdm}{\begin{equation*}}
\newcommand{\edm}{\end{equation*}}
\newcommand{\bea}{\begin{eqnarray}}
\newcommand{\eea}{\end{eqnarray}}

\newcommand{\partialf}[2]
{
 \ifthenelse{\equal{#1}{}}{\frac{\partial}{\partial #2}}{\frac{\partial #1}{\partial #2}}
}

\renewcommand{\(}{\left(}
\renewcommand{\)}{\right)}
\renewcommand{\[}{\left[}
\renewcommand{\]}{\right]}

\newcommand{\Del}{\Delta}

\newcommand{\df}{\textrm{d}}

\newcommand{\la}{\lambda}

\renewcommand{\b}{\beta}

\newcommand{\ii}{{\rm i}}

\newcounter{saveeqn}%

\newcommand{\defn}{\ensuremath{\stackrel{\mathrm{def}}{=}}}
\renewcommand{\equiv}{\defn}

\renewcommand{\U}{\mathbf{U}}

\begin{document}

\title{Statistical state dynamics based theory for the formation and equilibration of Saturn's north polar jet}


\author{Brian F. Farrell}
\affiliation{School of Engineering and Applied Science, Harvard University}
\author{Petros J. Ioannou}
\email{pjioannou@phys.uoa.gr}
\affiliation{Department of Physics, National and Kapodistrian University of Athens}
%

\date{\today}

\begin{abstract}
Coherent jets containing most of the kinetic energy of the flow are a common feature in observations of atmospheric turbulence at planetary scale. In the gaseous planets these jets are embedded in a field of incoherent turbulence on scales small relative to the jet scale. Large scale coherent waves are sometimes observed to coexist with the coherent jets and the incoherent turbulence with a prominent example of this phenomenon being the distortion of Saturn's North Polar Jet (NPJ) into a distinct hexagonal form. Observations of this large scale jet-wave-turbulence coexistence regime raises the question of identifying the mechanism responsible for forming and maintaining this turbulent state. The coherent planetary scale jet component of the turbulence arises and is maintained by interaction with the incoherent small-scale turbulence component. It follows that theoretical understanding of the dynamics of the jet-wave-turbulence coexistence regime can be facilitated by employing a statistical state dynamics (SSD) model in which the interaction between coherent and incoherent components is explicitly represented. In this work, a two-layer beta-plane SSD model closed at second order is used to develop a theory that accounts for the structure and dynamics of the NPJ. An asymptotic analysis is performed of the SSD equilibrium in the weak jet damping limit that predicts a universal jet structure in agreement with observations of the NPJ. This asymptotic theory also predicts the wavenumber (six) of the prominent jet perturbation. Analysis of the jet-wave-turbulence regime dynamics using this SSD model reveals that jet formation is controlled by the effective value of $\beta$ and the required value of this parameter for correspondence with observation is obtained. As this is a robust prediction it is taken as an indirect observation of a deep poleward sloping stable layer beneath the NPJ. The slope required is obtained from observations of the magnitude of the zonal wind component of the NPJ. The amplitude of the wave six perturbation then allows identification of the effective turbulence excitation maintaining this combined structure. The observed jet structure is then predicted by the theory as is the wave six disturbance. The wave six perturbation, which is identified as the least stable mode of the equilibrated jet, is shown to be primarily responsible for equilibrating the jet with the observed structure and amplitude.
\end{abstract}


\maketitle


\section{Introduction}

Coherent structures  emergent from small scale turbulence
are often observed in planetary atmospheres with the  zonal jets of the gaseous planets being familiar examples~\citep{Ingersoll-90, Vasavada-and-Showman-05, Sanchez-etal-2000,Galperin-etal-2014}.
While this phenomenon of spontaneous large scale jet organization from small scale turbulence has been extensively investigated in both observational and theoretical studies~\citep{Kraichnan-1967,Rhines-1975,Williams-79a, Williams-03,Panetta-93, Nozawa-and-Yoden-97,  Huang-Robinson-98, Lee-05, Manfroi-Young-99, Vallis-Maltrud-93,
Cho-Polvani-1996, Read-etal-2004,Showman-2007, Scott-Polvani-2008, Galperin-etal-04}  the physical mechanism underlying it remains controversial.
The prominence of jets in planetary turbulence  is in part due to the jet being a nonlinear stationary solution of the dynamics in the limit of vanishing dissipation and therefore not  disrupted  by nonlinear advection on the time scale of the large scale shear.
However, its being a stationary solution is insufficient by itself to serve as an explanation for the observed  jets  for three reasons.  First,  strong jets  typically assume a characteristic structure for a given set of system parameters, while any zonally symmetric flow is a fixed point of the inviscid dynamics.  Second,  nonlinear stationary states lack a mechanism of maintenance  against dissipation and so can not explain the fact that  the observed jets,  which are not maintained by  coherent external forcing such as by an imposed pressure gradient, persist much longer than the dissipation time scale.   Third, planetary  jets commonly appear to be unstable;  for example, the north polar jet (NPJ) of Saturn robustly satisfies the Rayleigh--Kuo
necessary condition for barotropic instability in a dissipationless stationary flow~\citep{Antunano-etal-2015, Sanchez-etal-2014}, and  barotropic instability of this jet has been verified by eigenanalysis~\citep{Barbosa-etal-2010}.


The aforementioned  considerations imply that a comprehensive theory for the existence of large scale jets in the atmospheres of the gaseous planets and in particular Saturn's NPJ must provide a mechanism for the formation and maintenance of the jet from incoherent turbulence,  the particular structure assumed by the jet and its stability.  In addition to these the
case of the NPJ also requires that the theory account for the prominent coherent wave six
perturbation that distorts the jet into a distinct hexagonal form.

The primary mechanism by which the large scale jets of the gaseous planets are maintained
is upgradient  momentum flux resulting from straining of the perturbation field by the mean jet shear  which
produces a spectrally nonlocal interaction between the small-scale perturbation field and the large-scale jet.
This mechanism has been verified in observational studies of both the
Jovian and Saturnian atmospheres~\citep{Ingersoll-etal-2004,Salyk-etal-2006, Delgenio-etal-2007},
as well as in numerical simulations~\citep{Nozawa-and-Yoden-97,Huang-Robinson-98,Showman-2007} and in laboratory experiments~\citep{Wordsworth-etal-2008}.
This upgradient momentum transfer mechanism has been found to maintain mean jets both in barotropic forced dissipative models~\citep{Huang-Robinson-98, Farrell-Ioannou-2003-structural} and in baroclinic free turbulence models~\citep{Panetta-93, Farrell-Ioannou-2009-closure} and can be traced to the interaction of the perturbation field with the mean shear
\citep{Farrell-Ioannou-1993-unbd, Huang-Robinson-98, Bakas-Ioannou-2013-jas,Srinivasan-Young-2014}.
Excitation of the observed small scale forced turbulence in the case of  both the Jovian and Saturnian jets
is believed to be of
convective origin ~\citep{Vasavada-etal-2006, Showman-2007,Gierasch-etal-2000, Sanchez-etal-2000,Porco-etal-2003,Delgenio-etal-2007,Showman-2007}. For our purposes  it suffices to maintain  the observed amplitude of  small scale field of turbulence.
We choose to maintain this turbulent field in the simplest manner though introducing a stochastic excitation. The structure
of the stochastic excitations is not important so long as it maintains the  observed amplitude of turbulence
given that  the anisotropy of the turbulence
is induced by the mean shear of the jet.

In this work Saturn's NPJ is studied using  the statistical state dynamics (SSD)
of a two-layer baroclinic model, specifically a closure at second order in its cumulant expansion  (cf.~Ref.~\citep{Hopf-1952}).
The implementation of SSD used is referred  to as the stochastic structural stability theory (S3T) system~\citep{Farrell-Ioannou-2003-structural}.   In S3T the nonlinear terms in the perturbation equation for the second cumulant
involves the third cumulant which is parameterized by a stochastic excitation rather than being explicitly calculated  while the nonlinear interaction of the perturbations with the mean jet are fully retained.  For this reason the S3T system may be described as
quasi-linear  (QL) in accord with the fact that quasilinearity is a general attribute of second order closures \citep{Herring-1963}.
S3T has been applied previously to the problem of jet formation in barotropic turbulence~\citep{Farrell-Ioannou-2007-structure, Srinivasan-Young-2012,Tobias-Marston-2013,Parker-Krommes-2013-generation,Bakas-Ioannou-2013-prl,Constantinou-etal-2014}
to jet dynamics in the shallow water equations  \citep{Farrell-Ioannou-2009-equatorial}
and to jet dynamics  in baroclinic turbulence~\citep{Farrell-Ioannou-2008-baroclinic, Farrell-Ioannou-2009-closure}. The S3T system employs an equivalently infinite ensemble in the dynamical equation for the second cumulant and as a result provides  an autonomous and fluctuation-free dynamics for the statistical mean turbulent state which greatly facilitates analytical study\footnote{
Formal    justification of the second order S3T closure has been given by Bouchet et al. \cite{Bouchet-etal-2013,
Bouchet-etal-2014} who
show that for $\alpha = \lambda \tau\ll1$ to leading order in $\alpha$ the statical dynamics
asymptotically approaches the dynamics of the S3T second-order closure  with
 mean flow $O(1/\alpha)$ larger than the perturbation field. The dimensionless parameter $\alpha$ is the product of
 $\tau= L_y/(U_{\rm max}-U_{\rm min})$,   the shear time of the jet,   and, $\lambda$, the inverse of  the time scale on which
 the large scale flow evolves,
which is inversely proportional to the square root of the energy injection rate, $\varepsilon$. However,
 these formal results are too conservative. For example, it has been demonstrated  that S3T theory
  is  predictive of the bifurcation
of the statistically homogeneous state of barotropic beta plane turbulence to the jet state, in which arbitrarily small jets emerge, and that these predictions are
valid  at  the bifurcation point, which is the regime of $\alpha \gg1$ \citep{Constantinou-etal-2014}.
Similar validity of perturbative structure instability in the  $\alpha>>1$ regime
has been demonstrated for the case of 3-D Couette flow turbulence  \citep{Farrell-Ioannou-2016-bifur}.
In retrospect we understand that the fundamental underlying reason for the robust validity of the S3T dynamics has a physical basis in that it captures the mechanism  determining the statistical state of shear
turbulence which is interaction between the mean flow and the perturbations supporting it  by means of
quadratic fluxes which are accurately obtained from the second cumulant. The fact that this interaction is
contained in the closure at second order is consistent with the success of this closure in capturing the
qualitative and in many examples also the quantitative behavior of turbulent equilibria in shear flow.}.

When applying  S3T  to the study of zonal jets it is useful to equate the ensemble mean and
zonal mean by appeal to the ergodic hypothesis.   A two-layer model is employed  in order to provide the possibility for
baroclinic and barotropic dynamics  both  for the jet itself and for the perturbations that are  involved in the  equilibration
dynamics.  One reason this is important is that
a barotropic, equivalent barotropic or shallow water model with the observed
Rossby radius would not allow the problem freedom to adopt
 barotropic dynamics corresponding to  formation of deep jets.
 In the event we find that the statistical equilibrium jets are either barotropic or close to it
so that the Rossby radius is not a relevant parameter \citep{Showman-2007}.

We find that  jet formation is tightly controlled by the effective vorticity gradient, $\beta$.    As this is a robust requirement of the dynamics, the observed jet structure  is taken as an indirect observation of this parameter.   Saturn's NPJ is similar  in structure and amplitude to strong midlatitude jets on the gaseous planets such as Jupiter's $24^\circ\;\rm N$ jet while the planetary value of
$\beta_{sat}(74^\circ)=1.6\times 10^{-12}~\rm m^{-1}\,s^{-1}$ at the latitude of the NPJ is too weak to stabilize a jet with the observed amplitude
($98.7~\rm m\,s^{-1}$),  which poses a
dynamical dilemma~\cite{Antunano-etal-2015}.   Theory and observation can be brought into correspondence by
inferring a deep  strongly statically stable layer beneath the jet giving rise to the equivalent of a topographic
$\beta$ effect.  The $\beta$ used in the model is then the
dyamical superposition of the effects of both the
planetary and the topographic components.   With this inferred effective $\beta$ the observed
jet structure accords with the theory.

While the first cumulant provides the structure of the jet, the second determines the planetary scale wave disturbance superposed on the jet.   With the inferred value of $\beta$ and an incoherent turbulence excitation level consistent with observation this wave is found to have wavenumber six and the amplitude required to produce the observed hexagonal shape of the NPJ.
The role of this wave in the dynamics is  to equilibrate the jet with the observed velocity structure and amplitude
while providing the pathway for  dissipation of the energy that the jet is continuously extracting from the small scale turbulence.

Previously advanced explanations for the prominent wave six perturbation to Saturns's NPJ are that it arises as the surface expression of an upward propagating Rossby wave the origin of which is attributed to a wave six  corrugation of an inferred deep lower layer \citep{
Sanchez-etal-2014} and that the wave six results from nonlinear equilibration of a linear instability of the jet
\citep{Morales-etal-2011,Morales-etal-2015}. However, the equilibrated wave six instability predicts
closed vortices which are not seen in observations of the NPJ.

The nonlinear S3T equilibrium obtained satisfies the Rayleigh--Kuo necessary condition for
barotropic jet instability in both the prograde and retrograde jet and significant interaction between the jet
and the modes associated with both these vorticity gradient structures is seen.  Although the
Rayleigh--Kuo
criterion is not sufficient to ensure instability of a barotropic jet, experience has shown that,  absent careful
contrivance of the velocity profile,  satisfaction of this necessary condition coincides  with
modal instability.  Therefore, finding this criterion satisfied absent an instability directs attention toward the mechanism
responsible for the implied careful contrivance.
In the case of the NPJ this mechanism  is shown in this work to be continuous feedback regulation between the coherent jet
and the incoherent turbulence that adjusts the jet to marginal stability  under conditions of sufficiently strong forcing by the
small scale incoherent components  to produce an unstable jet profile.
The widely debated enigma of  the stability of the zonal jets of the gaseous planets and in particular the stability of the
NPJ in the face of observed strong vorticity gradient  sign reversals is in this way
resolved by the jets having been adjusted to (in most cases marginal)
stability by  perturbation-mean flow interaction between the
first and second cumulants of the S3T dynamics.  This mechanism of regulation to marginal
stability by feedback
between the first and second cumulant is familiar as the agent underlying establishment of
turbulent  equilibria in Rayleigh-Benard convection \citep{Malkus-1956,Herring-1963} and in establishing the
baroclinic adjustment state in baroclinic turbulence \citep{Stone-1978,Farrell-Ioannou-2008-baroclinic,Farrell-Ioannou-2009-closure}.

This mechanism of equilibration  also has implications for
the problem of identifying how energy transferred upscale from the excited
small scales to large scales in geostrophic turbulence is
dissipated  as is required to maintain statistical
equilibrium.  Ekman damping associated
with no slip boundaries is not available in the absence of
solid boundaries and there is negligible diffusive damping
at the jet scale.  In fact, in the case of the NPJ, the eddy
fluxes, including the eddy damping, are explicitly
calculated for the SSD equilibrium state and these are found to be  dominantly upgradient and therefore in toto are responsible
for maintaining rather than dissipating the jets.  In model studies hypodiffusion
is commonly used to allow establishment of a
statistically steady state \citep{Danilov-04,Scott-Polvani-2007}.
While hypodiffusion is often employed  without physical justification it can be related to radiative damping of baroclinic structures
\citep{Scott-Polvani-2007}.
However, we find that the dynamics of jet formation result in primarily
barotropic jet structure so that thermal damping is not relevant.
Instead, in the case of Saturn's NPJ we identify the physical dissipation mechanism
responsible for equilibrating the jet
to be energy transfer directly from the coherent
jet to a wave six structure followed by dissipation of the
energy by this wave. We note
that in this planetary scale turbulence regime both the
upscale energy transfer maintaining the jet as well as the
downscale energy transfer to the wave six mode regulating
its amplitude occur directly between remote scales and
in neither case do these involve a turbulent cascade.

\section{Applying S3T to study the SSD equilibria in a two-layer model of Saturn's atmosphere}

We wish to choose a model configured to address the question of whether the planetary jet
structure is deep or shallow; that is, whether it is confined to a shallow surface layer and therefore favors dynamically a
baroclinic structure  or if the dynamics favors establishment of a deep barotropic structure.
The simplest model that retains the freedom for the dynamics to exploit both baroclinic and barotropic processes and to attain either
baroclinic or barotropic structure for the coherent component of the turbulent state equilibria
is the  quasi-geostrophic
two-layer model.
We choose parameters appropriate for the NPJ of Saturn including  planetary
vorticity gradient $\beta = \df f_0/ \df y $, where $f_0$ is the planetary vorticity and the derivative
is taken at the center of the channel at $74^\circ\;\rm N$.  The  channel  size is $L_x$ in the zonal, $x$, direction, and $L_y$
in the meridional direction, $y =R \phi$, in which $R$ is the radius of the planet and $\phi$ is the latitude.
The layers are of equal depth, $H$,
with the density of the upper layer, $\rho_1$,  being less than that of the lower, $\rho_2$.
The stream function in each layer is  denoted $\psi_j$, with $j=1$
referring to the top layer and $j=2$ to the bottom
layer.  The zonal velocities  are $u_j= -\partial_y \psi_j$ and the meridional velocities are $v_j = \partial_x \psi_j$ ($j=1,2$).
The  dynamics is expressed as  conservation of potential vorticity,  $q_j = \Del \psi_j +\beta y +(-1)^j  \lambda^2(\psi_1-\psi_2)$),
in which $\lambda^2= f_0^2/(g' H) $,
with $g'=g (\rho_2-\rho_1)/{\rho_0}$ the reduced gravity associated with the planetary gravitational acceleration $g$
and $\rho_0$ is a characteristic density of the fluid, which is taken here to be $(\rho_1+\rho_2)/2$ (cf.~Ref.~\citep{Pedlosky-1992}).  The Rossby radius of deformation for baroclinic
motions in this two-layer fluid
is $L_d = 1/(\sqrt{2}\lambda)$.

The quasi-geostrophic dynamics expressed  in terms of the barotropic  $\psi = ( \psi_1+\psi_2)/2$ and
baroclinic $\theta = (\psi_1-\psi_2)/2$ streamfunctions is:
\begin{subequations}
\label{eq:BRCL}
\begin{align}
\partial_t \Del \psi &+ J(\psi,\Del\psi) + J(\theta,\Del\theta) + \beta \partial_x\psi  =\nonumber \\
& \hspace{9em}= -r \Del \psi  +\sqrt{\varepsilon}\,f_\psi~,\\
\partial_t \Del_\la \theta &+ J(\psi,\Del_\lambda\theta) + J(\theta,\Del\psi) + \beta \partial_x\theta =\nonumber \\
& \hspace{9em}=-r \Del_\la \theta +\sqrt{\varepsilon}\,f_\theta~,
\end{align}\end{subequations}
where $\Del_\lambda\equiv\Del -2\lambda^2$. Terms $f_\psi$ and $f_\theta$ are  random functions with zero mean
representing independent  vorticity excitations of the fluid by unresolved processes,  like  convection,
with amplitude controlled by $\varepsilon$.
The advection of potential vorticity is expressed using the
Jacobian $J(f,g)=(\partial_x f)(\partial_y g) - (\partial_yf)(\partial_x g)$.
Equations~(\ref{eq:BRCL}) are  non-dimensional with length scale $L=1000~\rm km$ and time scale
$T= 1$  Earth day implying  velocity unit $11.5~\rm m\,s^{-1}$. The coefficient of linear damping is $r$ and this damping may vary with the scale of the motions when appropriate for probing the dynamics controlling jet formation and equilibration.  In particular,
 insight can be gained by examining  the regime in which the large scale zonal flow
is damped at a rate $r_m\ll r_p$ where $r_p$ is the  damping rate of the perturbations.
Relatively small damping rate for the large scale jet compared to the small scale incoherent
turbulence is expected on physical grounds
if the damping is diffusive so that the rate is proportional to total square wavenumber.
Radiative damping proportional to $\theta$
would correspond to  second order hypodiffusion  on the baroclinic component of the jet potential vorticity
but we find the jets are essentially barotropic
so that
radiative damping would be ineffective.
The NPJ is sufficiently lightly damped that its structure is determined primarily by nonlinear feedback
regulation between the jet and the incoherent component of the turbulence with a negligible role for
jet-scale damping in the equilibration process. Simplifying the problem by eliminating jet damping
altogether allows
study of a physically realistic asymptotic  regime  in which the equilibrium state is completely determined by
nonlinear feedback regulation.  We verify that inclusion of a small damping rate makes no substantive change
in the jet equilibrium obtained in the undamped jet limit.


The barotropic and baroclinic streamfunctions are decomposed into a zonal mean (denoted with capitals) and deviations from
the zonal mean (referred to as perturbations and denoted with primed small letters):
\begin{equation}
\psi=\Psi + \psi'~,~~\theta= \Theta+\theta'~.
\end{equation}
We denote the barotropic zonal mean flow as $U=-\partial_y \Psi$  and  the baroclinic zonal mean flow as $H=-\partial_y \Theta$.
Equations for the evolution of the barotropic and baroclinic zonal mean flows are obtained by formimg the zonal mean of~\eqref{eq:BRCL}:\begin{subequations}\label{eq:NL_mbrcl}
\begin{align}
\partial_t U &=  \overline{v' q'}_{\psi} -r_m U\ , \\
\partial_t D^2_\la H  &=D^2 \overline{v' q'}_{\theta}-r_m D^2_\la H\ ,
\end{align}\end{subequations}
in which the overline denotes zonal averaging,
$D^2 \equiv \partial_y^2$, $D^2_\la \equiv D^2-2\la^2$, and $r_m$ denotes the linear damping rate of the mean flow. The terms
\begin{subequations}\begin{align}
\overline{v' q'}_{\psi}&\equiv \overline{ (\partial_x\psi')(\partial^2_y \psi') + (\partial_x\theta') (\partial^2_y\theta')}\ ,\\
\overline{v' q'}_{\theta}&\equiv \overline{(\partial_x\psi') D^2_\la \theta' +  (\partial_x\theta')(\partial^2_y\psi')}\ ,\end{align}\end{subequations} \noindent
are, respectively, the Reynolds stress divergence forcing of the barotropic and baroclinic mean flow or  equivalently the barotropic and baroclinic vorticity flux.
Vorticity  fluxes are referred to as upgradient if they
have the tendency to reinforce the  mean flow, so that e.g. $\int_0^{L_y} U\,\overline{{v' q'}}_{\psi} \;\df y>0$;
otherwise they are termed downgradient.

The evolution equations for the perturbations are:
\begin{subequations}
\label{eq:NL_pbrcl}
\begin{align}
&\partial_t \Del \psi' + U  \partial_x\Del\psi' + H \partial_x\Del\theta' + (\beta - D^2 U ) \partial_x \psi'  \nonumber \\
&~  - D^2 H  \partial_x \theta'=  -r_p \Del \psi'  
- J(\psi',\Del \psi' )' -J (\theta', \Del \theta')' ,\label{eq:psi'}\\
&\partial_t \Del_\la \theta' + H  \partial_x\Del\psi' + U \partial_x\Del_\la \theta'
+ (\b - D^2 U) \partial_x \theta' \nonumber\\
&
~- D^2_\la H \partial_x \psi' =
 -r_p \Del_\la \theta' 
- J(\psi',\Del_\la \theta' )' -J (\theta', \Del \psi')'~,\label{eq:theta'}
\end{align}
\end{subequations}
with the prime Jacobians denoting the perturbation-perturbation interactions,
\begin{equation}
J(f,g)'=J(f,g)-\overline{J(f,g)}~.
\end{equation}
Equations~(\ref{eq:NL_mbrcl}) and~(\ref{eq:NL_pbrcl}) comprise the non-linear  system (NL) that governs the two layer baroclinic flow.
As previously remarked, dissipation of the mean at rate $r_m$
and of the perturbations at a lower rate, $r_p$,  in~(\ref{eq:NL_mbrcl}) and~(\ref{eq:NL_pbrcl}) is consistent with parameterizing diffusion while retaining the simplicity of linear damping.   More importantly, it allows us to explore the dynamically interesting regime $r_m=0$ in which the equilibration of the jet by nonlinear interaction with the perturbations is independent of jet damping.

We impose periodic boundary conditions  on $\Psi$, $\Theta$, $\psi'$, $\theta'$  at the channel northern and southern boundaries~\cite{Haidvogel-Held-1980, Panetta-93}.
These boundary conditions can be verified to require  
that the temperature difference between the channel walls
remains fixed. In this work we have chosen to isolate the primarily barotropic nature of jet formation and equilibration dynamics by taking this temperature difference to be zero. Simulations including baroclinic influences arising from temperature gradients below the threshold required for baroclinic instability, as is appropriate for Saturn, show small changes in the results \cite{Farrell-Ioannou-2008-baroclinic,Farrell-Ioannou-2009-closure}.

The  corresponding quasi-linear system (QL) is obtained by substituting for the perturbation--perturbation interactions in~\eqref{eq:NL_pbrcl} a state independent and temporally delta correlated stochastic excitation together with sufficient added
dissipation to obtain an approximately energy conserving closure~\citep{DelSole-Farrell-1996,DelSole-Hou-1999,DelSole-04}. Under these assumptions the QL perturbation equations in matrix form for the Fourier components of the barotropic and baroclinic streamfunction are:
\begin{subequations}
\label{eq:QLbrcl}
\begin{align}
\frac{\df\psi_k}{\df t} &= \A_k^{\psi \psi} \psi_k + \A_k^{\psi \theta} \theta_k + \sqrt{\varepsilon} \,\Delta_k^{-1} \F_k\,\xi^{\psi} (t) ~,\\
\frac{\df\theta_k}{\df t} &= \A_k^{\theta \psi} \psi_k + \A_k^{\theta \theta} \theta_k + \sqrt{\varepsilon} \,\Delta_{k \la}^{-1} \F_k \,\xi^{\theta} (t) ~.
\end{align}
\end{subequations}
The variables in~\eqref{eq:QLbrcl} are the Fourier components of the perturbations fields defined through e.g.,
\be
\psi'(x,y_i,t)=  \sum_{k>0}\Re \[ \psi_{k,i}(t)\,e^{\ii k x} \]\ ,
\ee
with $\Re$ denoting the real part.
The states $\psi_k$ and $\theta_k$ are column vectors with entries the complex value of the barotropic and baroclinic streamfunction at the collocation points $y_i$. The excitations, which represent both the explicit excitation and the stochastic parameterization of the perturbation--perturbation interactions in the perturbation equations, are similarly expanded so that the excitation at collocation point $y_i$ is given through e.g.,
\be
 f_\psi (x,y_i,t) =   \sum_{k>0}\sum_j\Re \[ F_{k, i j} \xi_j^\psi(t)\,e^{\ii k x} \]\ .
\ee
Terms $\xi^{\psi}$ and $\xi^{\theta}$ are independent temporally delta-correlated  complex vector stochastic processes with zero mean satisfying:
\begin{subequations}
 \begin{gather}
\langle  \xi^\psi (t_1)\xi^{\psi \dagger}(t_2) \rangle = \langle \xi^\theta (t_1)\xi^{\theta \dagger}(t_2) \rangle =  \delta(t_1-t_2)\,\I~,\\
\langle \xi^\psi (t_1)\xi^{\theta \dagger}(t_2) \rangle = 0~,
\end{gather}\end{subequations}
in which $\langle\;\boldsymbol{\cdot}\;\rangle$ denotes the ensemble average over  forcing realizations, $\I$ the identity matrix and
$\dagger$ the Hermitian transpose.
 This excitation  is homogeneous  in the zonal  direction and identical  in each layer. In order
to ensure homogeneity  in $y$  the latitudinal structure  matrices $F_k$  are chosen so that their $(i,j)$ entry is a function of $|y_i-y_j|$.
The operators $\A_k$ that depend on  the mean flow
${\cal U} \equiv  [U,H]$
have components:
\begin{equation}
\A_k({\cal U})~=~
\left(%
\begin{array}{cc}
\A_k^{\psi \psi} & \A_k^{\psi \theta} \\
 \A_k^{\theta \psi} & \A_k^{\theta \theta} \\
\end{array}%
\right)\ ,
\end{equation}
with entries:
 \begin{subequations}
\label{eq:Abrcl}
\begin{align}
 \A_k^{\psi \psi}&=\Delta_k^{-1} \left [ -\ii k  U
 \Delta_k -\ii k \left (
\beta - D^2 U \right ) \right ]  -r_p~+\nu \Delta_k~, \\
 \A_k^{\psi \theta}&= \Delta_k^{-1} \left [ -\ii k  H \Delta_{k } +\ii k D^2 H
\right ]~, \\
 \A_k^{\theta \psi}& = \Delta_{k \la}^{-1} \left [ -\ii k  H \Delta_k
 +\ii k D^2_\la H  \right ] ~, \\
 \A_k^{\theta \theta}& = \Delta_{k \la}^{-1} \left [ -\ii k U \Delta_{k \la}
 -\ii k \left ( \beta - D^2 U \right ) 
 \vphantom{\left( \beta - D^2 U \right )} +\nu \Delta_k \Delta_k \right ]-r_p~,
\end{align}\end{subequations}
in which $\Delta_k \equiv D^2- k^2$, $\Delta_{k \la} \equiv \Delta_k - 2 \lambda^2$.
Diffusion  is included  in the perturbation dynamics  for numerical stability and its coefficient, $\nu$, is  set equal to the square of the grid interval.

The corresponding S3T statistical state dynamics  system expresses the dynamics of an equivalently infinite ensemble of realization of  the  QL equations~(\ref{eq:QLbrcl}),
with each ensemble member  sharing the same mean ${\cal U}$ while being excited by an independent noise process,.  This ensemble of perturbation equations is coupled to the mean equation~\eqref{eq:NL_pbrcl} through the ensemble  mean vorticity fluxes:
$\langle \overline{v' q'}_\psi \rangle$
and  $\langle \overline{v' q'}_\theta \rangle$.  Identification of
the  ensemble mean  with the zonal mean is made by appeal to the ergodic hypothesis
(cf.~Ref.~\cite{Farrell-Ioannou-2003-structural}).
The appropriate perturbation variable for this SSD is  the covariance matrix, which is the second cumulant of the statistical state dynamics.
The covariance for the wavenumber $k$ zonal Fourier component is defined as:
\begin{equation}
\C_k ~= ~
\left(%
\begin{array}{cc}
\C_k^{\psi \psi} & \C_k^{\psi \theta} \\
\C_k^{\psi \theta \dagger} & \C_k^{\theta \theta} \\
\end{array}%
\right)~,
\end{equation}
where $\C_k^{\psi \psi}= \langle \psi_k \psi_k^{ \dagger}\rangle $,
 $\C_k^{\psi \theta}= \langle \psi_k \theta_k^{\dagger} \rangle$,
 $\C_k^{\theta \theta }= \langle \theta_k \theta_k^{\dagger} \rangle $.
 The ensemble mean vorticity fluxes in~\eqref{eq:NL_mbrcl}  are expressed in terms of the  SSD perturbation variable $\C_k$ as:
\begin{subequations}
\label{eq:vq}
\begin{align}
\langle  \overline{v' q'}_\psi \rangle &\equiv \sum_k  \langle \overline{v' q'}_{\psi} \rangle_k  \nonumber \\
&
 =\sum_k \frac{k}{2}\,{\rm diag} \left [ \Im \( D^2 \C_k^{\psi \psi} + D^2 \C_k^{\theta \theta} \)\right ] ~,
\label{eq:vqpsi}\end{align}\begin{align}
 \langle  \overline{v' q'}_\theta \rangle & \equiv \sum_k  \langle \overline{v' q'}_{\theta} \rangle_k  \nonumber \\
& = \sum_k \frac{k}{2}\,{\rm diag} \left [ \Im \left (
 D^2_\lambda \C_k^{\psi \theta \dagger} + D^2 \C_k^{\psi \theta} \right ) \right ] ~,
 \label{eq:vqtheta}
\end{align}
\end{subequations}
 with  the $\rm diag$ operator selecting the diagonal elements of a matrix and $\Im$ denoting the imaginary part.
 The fluxes are evaluated at each time  from  $\C_k$ as it evolves  according to the Lyapunov equation:
\begin{equation}
 \frac{\df \C_k }{\df t} =\A_k({\cal U}) \, \C_k + \C_k \,\A_k^{\dagger}({\cal U}) + \varepsilon \Q_k~,\label{eq:ensemblea}
\end{equation}
with $\Q_k$ the covariance of the stochastic excitation (cf.~Ref.~\citep{Farrell-Ioannou-2003-structural,Farrell-Ioannou-2008-baroclinic}). The covariances $\Q_k$ are normalized so that for each  $k$  an equal amount of  energy is injected per unit time so that the excitation rate  is  controlled by the parameter $\varepsilon$. The normalization is chosen so that $\varepsilon=1$ corresponds to injection of $10^{-4}~\rm W\,m^{-2}\,kg^{-1}$. Note that because the excitation has been assumed temporally delta-correlated
this energy injection rate  is independent of the state of the system.

We consider two types of excitation. When both layers are  independently excited (this case is indicated with $E_{1,2}$) the covariance of the excitation is:
\begin{equation}
\Q_k~=~
\left(%
\begin{array}{cc}
\Delta_k^{-1} \F_k \F_k^{\dagger} \Delta_k^{-1 \dagger}& 0 \\
 0 & \Delta_{k \la}^{-1} \F_k \F_k^\dagger\Delta_{k \la}^{-1 \dagger}\end{array}%
\right)~.
\end{equation}

When only the top layer is excited (case indicated $E_1$) the covariance is given by:\begin{equation}
\Q_k~=~
\left(%
\begin{array}{cc}
\Delta_k^{-1} \F_k \F_k^{\dagger} \Delta_k^{-1 \dagger}& \Delta_k^{-1} \F_k \F_k^{\dagger} \Delta_k^{-1 \dagger} \\
 \Delta_k^{-1} \F_k \F_k^{\dagger} \Delta_k^{-1 \dagger}& \Delta_k^{-1} \F_k \F_k^{\dagger} \Delta_k^{-1 \dagger}
\end{array}%
\right)~.
\end{equation}

\begin{figure}
	\centering
	\includegraphics[width = \columnwidth]{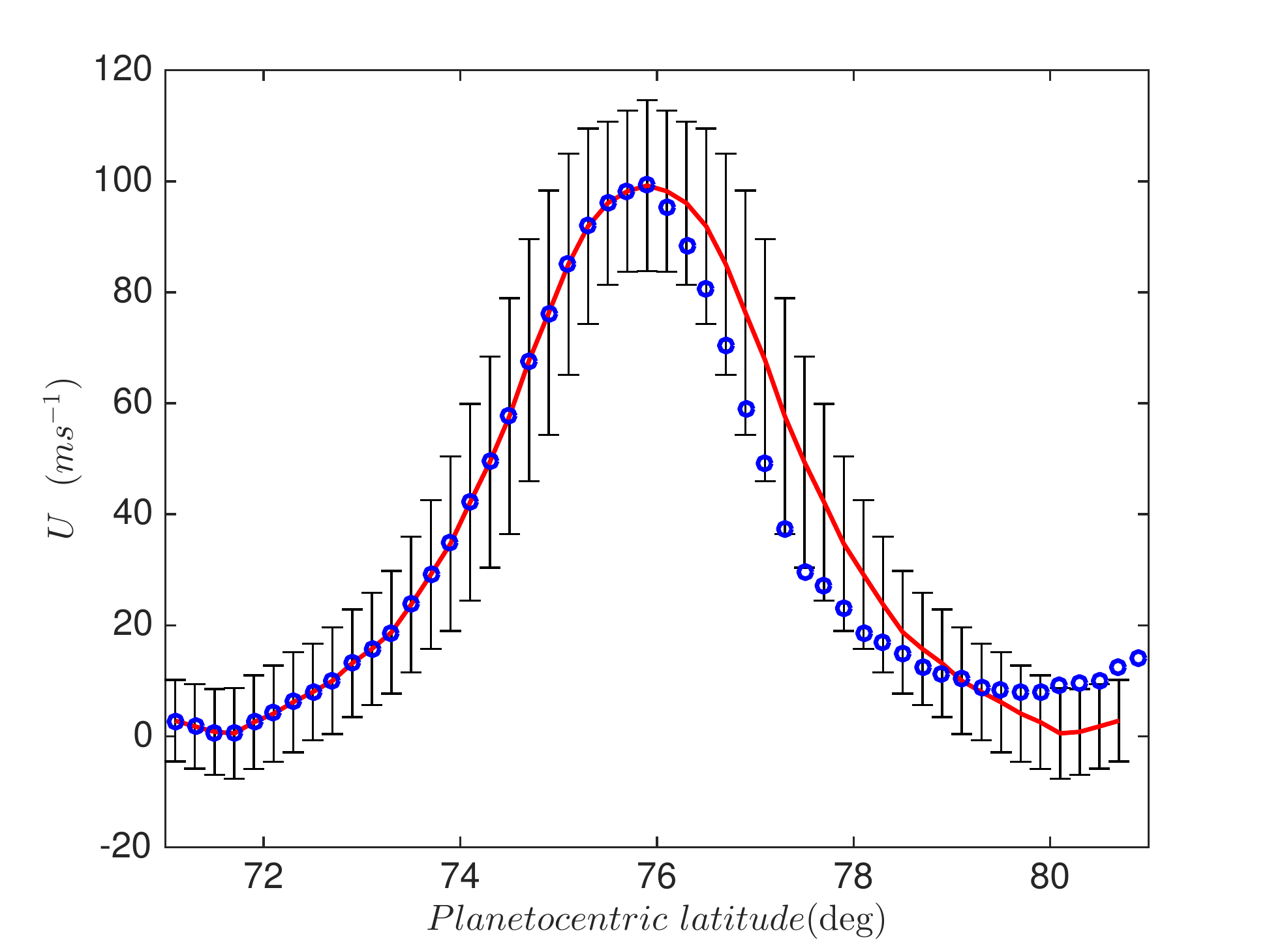}
	\caption{Observed NPJ velocity (circles) from~Ref.~\citep{Antunano-etal-2015}. The observed jet has been symmetrized by reflecting its southern flank about the jet maximum (solid line with one standard deviation error bounds).}
\label{fig:U}
 \end{figure}

The S3T  dynamics for  the evolution of the first two cumulants of the flow takes the form:
\begin{subequations}
\label{eq:S3T}
\begin{align}
\frac{\df U}{\df t} &= \sum_k~ \langle \overline{v' q'}_\psi \rangle_k -r_m U ~,\label{eq:mean1}\\
 \frac{\df H}{\df t} &= D^{-2}_\lambda D^2 \sum_k  \langle \overline{v' q'}_\theta \rangle_k
-r_m H~, \label{eq:mean2}\\
\frac{\df \C_k }{\df t}& =~\A_k({\cal U}) \, \C_k~+\C_k \,\A_k^{\dagger}({\cal U})~+~\varepsilon \Q_k~,\label{eq:pertC}\end{align}\end{subequations}
with the vorticity fluxes given in terms of the $\C_k$ in~\eqref{eq:vq} and the operators $\A_k$ defined in~\eqref{eq:Abrcl}.

The S3T system represents the  second cumulant with an infinite perturbation ensemble and it is as a result
autonomous and
therefore has the very useful
property for theoretical investigation of providing exact stationary fixed point solutions for statistical equilibrium states.  Because the excitation is  spatially homogeneous the zero mean flow,  $U=H=0$,   together with the perturbation field,
$\C_k^e$, satisfying the corresponding steady state Lyapunov equations, is an  equilibrium solution for
any $\varepsilon$~\eqref{eq:pertC} (for the explicit expression of the equilibrium covariance see Ref.~\cite{DelSole-Farrell-95}).
However,  this homogeneous equilibrium state is  unstable in the S3T system for $\varepsilon $ greater than a critical
$\varepsilon_c$.  This critical value of excitation rate resulting in unstable jet growth in the S3T system can be found by analyzing the stability of perturbations from this equilibrium state using the perturbation form of~\eqref{eq:S3T} (cf.~Ref.~\citep{Farrell-Ioannou-2008-baroclinic}).
As $\varepsilon$ is increased beyond $\varepsilon_c$ this instability results in a bifurcation in which finite amplitude jet equilibrium solutions,
${\cal U}^e= [U^e,H^e]$, emerge with associated covariances $\C_k^e$ that satisfy the time independent equilibrium state of
~\eqref{eq:S3T} for this  $\varepsilon$.   These equilibria
are stable for a range of $\varepsilon$ and satisfy the steady state equations:
\begin{subequations}\label{eq:S3Teq}\begin{gather}
\sum_k \langle \overline{v' q'}_\psi \rangle_k = r_m U^e ~,~~D^2 \sum_k \langle \overline{v' q'}_\theta \rangle_k = r_m D^2_\lambda H^e~,\label{eq:meane}\\
\A_k({\cal U}^e)\,\C_k^e + \C_k^e \,\A_k^{\dagger}({\cal U}^e) = -\varepsilon \Q_k~.\label{eq:pertCe}
\end{gather}\end{subequations}
Remarkably, these equilibria exist even for $r_m=0$. This limit is especially useful for theoretical investigation because the associated  equilibria have a universal form: when $r_m=0$ it follows
from the linearity of~\eqref{eq:pertCe} that if ${\cal U}^e$, $\C_k^e$  (for the excited $k$) is an equilibrium solution for $\varepsilon=1$ then  the same ${\cal U}^e$ is an equilibrium solution with
$\varepsilon C_k^e$ for any $\varepsilon$.
It does take longer to reach the equilibrium state with small $\varepsilon$ but the same equilibrium is eventually established by nonlinear
feedback regulation between the mean and perturbation equations.
However, it is important to note that for large enough $\varepsilon$ this equilibrium solution may itself become S3T unstable.

\section{Parameters}

The first 56 zonal wavenumbers, $k = 2 \pi n / L_x$ with $n=1,\dots,56$, are excited in the perturbation dynamics. These are referred to alternatively
as global wavenumbers or as waves $n=1,\dots,56$.
The simulations use
$64$ grid points in $y$ with convergence verified by doubling this resolution.
The stochastic excitation has  Gaussian structure in $y$  with
$\F_k $ chosen so that the $(i,j)$ element of the excitation is
proportional to $e^{-(y_i-y_j)^2/\delta^2}$, with $\delta=1$.
Recall that  the associated excitation covariances in the S3T dynamics, $\Q_k$, are normalized so that each wavenumber provides the same energy injection rate and that with  $\varepsilon=1$ the total  energy injection rate over  all wavenumbers is dimensionally $10^{-4}~\rm W\,kg^{-1} $.
We have chosen for modeling the NPJ
a doubly periodic channel with parameter values: $L_y = 10^4~\rm km$, $L_x=8\times 10^4~\rm km$,
$\beta_{sat}(74^\circ)=1.6\times 10^{-12}~\rm m^{-1}\,s^{-1}$,
$\lambda = 10^{-3}~\rm km^{-1}$, perturbation damping $r_p=0.2~\rm day^{-1}$ and excitation $\varepsilon=1$.

 \begin{figure*}
	\centering
       \includegraphics[width = .6\textwidth]{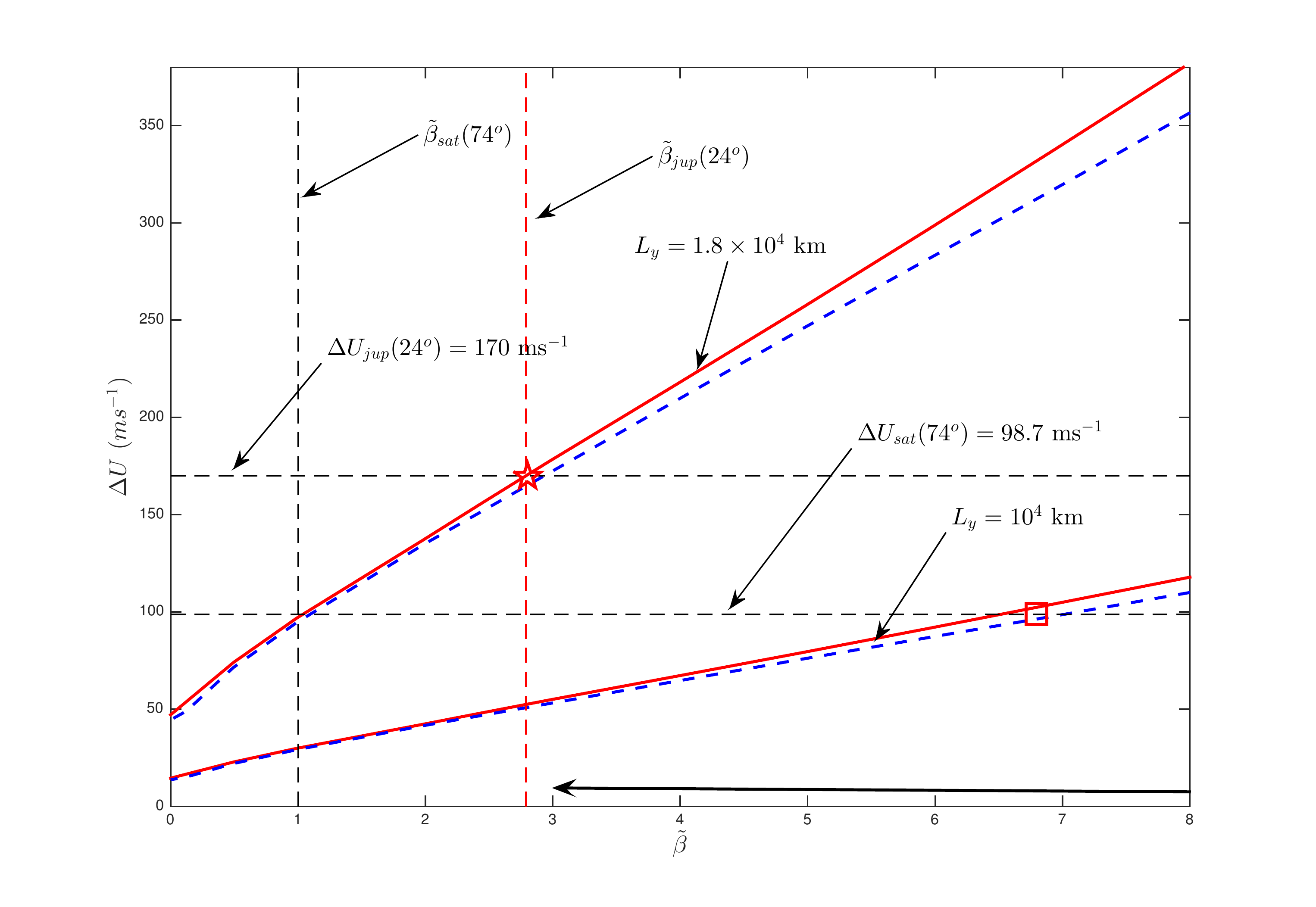}
        \caption{S3T equilibrium jet amplitude, $\Delta U$, as a function of  $\tilde \beta = \beta/ \beta_{sat}(74^\circ)$ in the low jet damping regime. $\Delta U$ is the difference between the maximum and the minimum jet velocities. Indicated are values of $\tilde \beta$ for Saturn's  NPJ ($L_y=10^4~\rm km$) and for Jupiter's  $24^\circ\;\rm N$ jet ($L_y=1.8\times 10^4~\rm km$). The equilibria for this case in which equal excitation has been imposed in both layers are barotropic and continuous lines indicate these equilibria for $r_m=0$.  Dashed lines showing these equilibria for $r_m = 0.001~\rm day^{-1}$ differ little indicating the validity of the $r_m=0$ asymptotic. The dependence on $\varepsilon$ and $r_p$ of these asymptotic jets  is also very weak and is not shown.
        The asymptotic scaling of the jet amplitude $\Delta U$ with $\beta L_y^2$ is nearly perfect
        for $\tilde \beta > 3$  (marked with an arrow).  
        Also indicated are the planetary values of $\tilde \beta_{sat}(74^\circ)$, $\tilde \beta_{jup}(24^\circ)$, the observed $\Delta U_{sat}(74^\circ) = 98.7~\rm m\,s^{-1}$ from Ref.~\citep{Antunano-etal-2015} and the observed $\Delta U_{jup}(24^\circ)=170~\rm m\,s^{-1}$ from Ref.~\cite{Sanchez-etal-2008}.
        While the observed $\Delta U_{jup}(24^\circ)$ (star) is consistent with the planetary value of $\beta$ as predicted by the asymptotic theory, the observed $\Delta U_{sat}(74^\circ) $ (square) requires $\tilde \beta=6.9$ for consistency.}
\label{fig:DeltaU}
 \end{figure*}

 \section{Universal structure and amplitude  scaling of weakly damped turbulent jets}

The velocity structure of the NPJ  is  asymmetric  presumably because the  jet is
influenced by  encroachment of the  polar vortex flow on its north side (cf. Fig.~\ref{fig:U}). For simplicity we model a symmetric channel and consistently  choose to compare our results with a symmetrized jet obtained by reflecting the southern half of the observed jet structure about the jet maximum. This symmetrized jet  together with one standard deviation error bounds as tabulated in~\citep{Antunano-etal-2015} is shown in Fig.~\ref{fig:U}.

We anticipate  that the jet dynamics  will  be in the small jet damping regime in which
the exact value of  $r_m$  is  irrelevant and can be taken to vanish. 
With the remaining parameter value  given above we find that a barotropic jet with a single maximum in zonal velocity arises as an unstable S3T eigenmode.
That the jet structure is barotropic
is an important prediction of the SSD that remains
valid at finite amplitude.  Whether the jets of the
gaseous planets are deep or shallow has implications for discriminating among
mechanisms for jet formation.
We conclude that the dynamics favors deep jets.  A further implication is that imposition
of a finite Rossby radius in a barotropic model would not  be physically justified.
Consistent  with the barotropic structure of the jets,
we have verified that
the Rossby radius has little affect on the jet dynamics.
We wish to study the influence on the S3T  jet and perturbation covariance of the perturbation
damping rate, $r_p$, the amplitude of the excitation, $\varepsilon$, and $\beta$.
In the limit of small $r_m$ the time required to reach equilibrium depends
on $r_m$ but the final equilibrium jet structure depends to a good approximation only on the channel width and $\beta$.
In this regime, the equilibrium jet amplitude and structure
are  nearly independent of the excitation rate, $\varepsilon$, and the perturbation damping rate, $r_p$,
while the perturbation energy is to a very good approximation proportional to  their ratio $\varepsilon/ r_p$.
The  jet structure obtained in the limit $r_m \rightarrow 0$ is close to the observed structure so we exploit the simplicity of this limit by studying the jet dynamics  with zero mean jet damping, $r_m=0$.  Departures from the $r_m=0$ equilibrium solution resulting from physically relevant nonzero jet damping rates are verified to be small (cf. Fig.~\ref{fig:DeltaU}).

\begin{figure*}
   \centering
   \includegraphics[width = .6\textwidth]{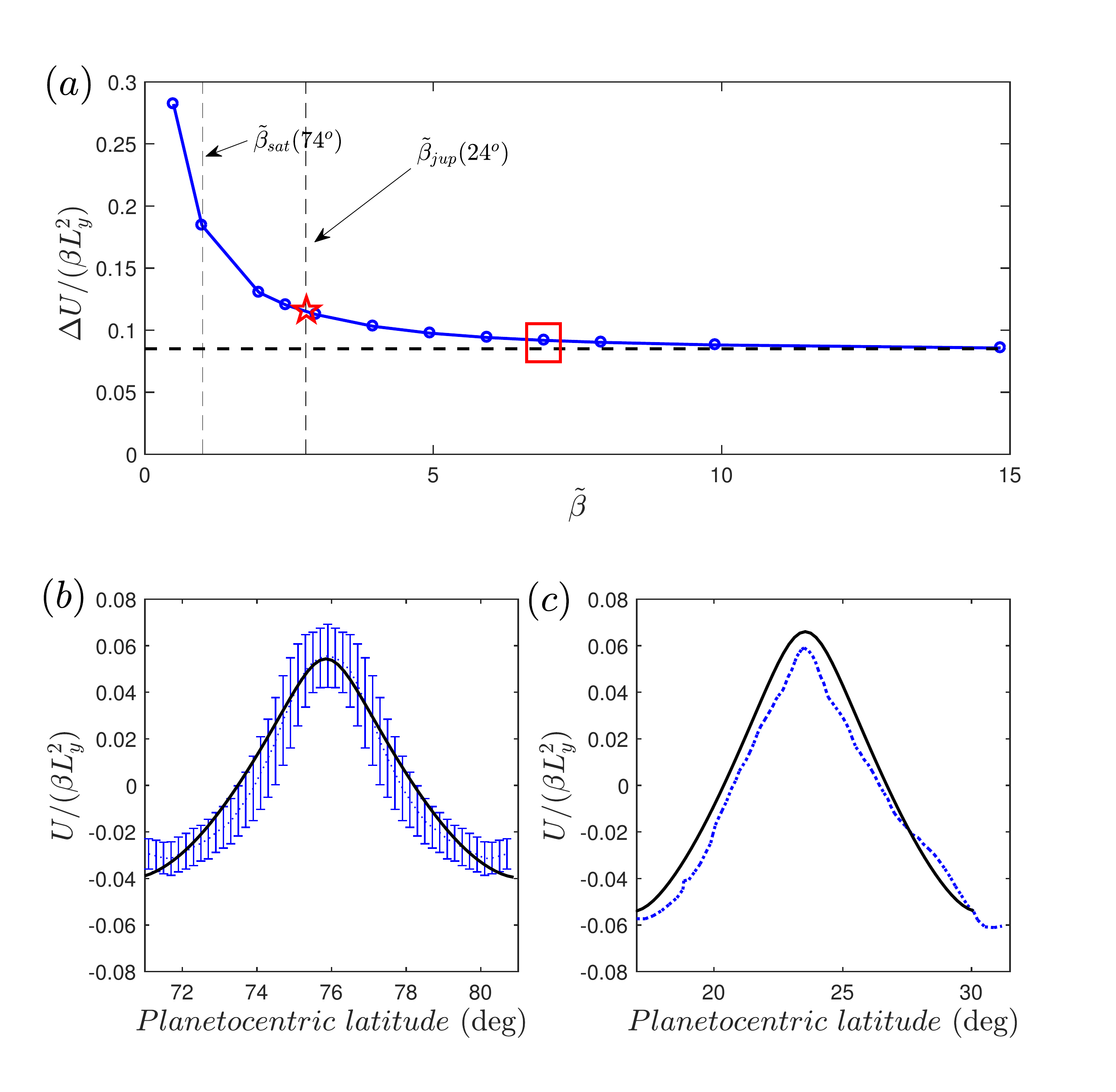}
   \caption{Universal structure of the  S3T equilibrium jets  for vanishing jet damping, $r_m=0$. Panel (a): normalized   equilibrium jet  amplitude, $\Delta U  /  (\beta L_y^2)$, as a function of $\tilde \beta = \beta/ \beta_{sat}(74^\circ)$.
        For large $\tilde \beta$ the S3T equilibrium flows assume an asymptotic structure and amplitude  $\Delta U  / ( \beta L_y^2) \approx 0.085$.
        Panel (b):   the observed NPJ  from Ref.~\citep{Antunano-etal-2015}  which has been symmetrized, scaled and had its  mean removed
        (dotted with error bounds)  compared to the scaled S3T equilibrium  jet for $\tilde \beta =6.9$ (indicated with a square in Panel (a)).
        Panel (c): the observed Jupiter $24^\circ\;\rm N$ jet from Ref.~\cite{Sanchez-etal-2008} with has been similarly symmetrized, scaled and had its
        mean removed (dotted) compared to the scaled S3T equilibrium jet for $\tilde \beta =2.8$ (indicated with a star in Panel (a)) for channel size $L_y=1.8 \times 10^4~\rm km$.
        These barotropic equilibria are obtained with  the two layers  equally excited with $\varepsilon =1$
        corresponding to energy injection of $10^{-4}~\rm W\,kg^{-1}$
        and $r_p=0.2~\rm day^{-1}$. This figure confirms that these planetary jets correspond to $r_m=0$ S3T equilibrium solution and approach the predicted asymptotic structure as $\tilde \beta$ increases.}
\label{fig:Unorm}
 \end{figure*}

With $r_m=0$ the equilibrium jet velocity  varies approximately linearly with $\beta L_y^2$ as shown in Fig.~\ref{fig:DeltaU} in which the values of $\beta$ appropriate for Saturn's NPJ and for Jupiter's $24^\circ\;\rm N$ jet are indicated. As $\beta$ increases the equilibrium jet assumes a universal
structure with this $\beta L_y^2$ scaling, as shown in Fig.~\ref{fig:Unorm}.
While Jupiter's $24^\circ$ N jet  corresponds closely with this universal scaling (cf. Fig.~\ref{fig:Unorm}c),
Saturn's NPJ is observed to be substantially stronger at $98.7~\rm m\,s^{-1}$ than the approximately $30~ \rm m\,s^{-1}$ (cf. Fig.~\ref{fig:DeltaU}) predicted by the scaling for
$\beta_{sat}(74^\circ)=1.6\times 10^{-12}~\rm m^{-1}\,s^{-1}$ and the NPJ channel width of $L_y = 10^4~\rm km$.  The effective value of $\beta$ required to obtain correspondence with the scaling is $\beta_{eff}=6.9 \beta_{sat}(74^\circ)$

\begin{figure*}
	\centering
       \includegraphics[width = .55\textwidth,trim = 0mm 22mm 0mm 0mm, clip]{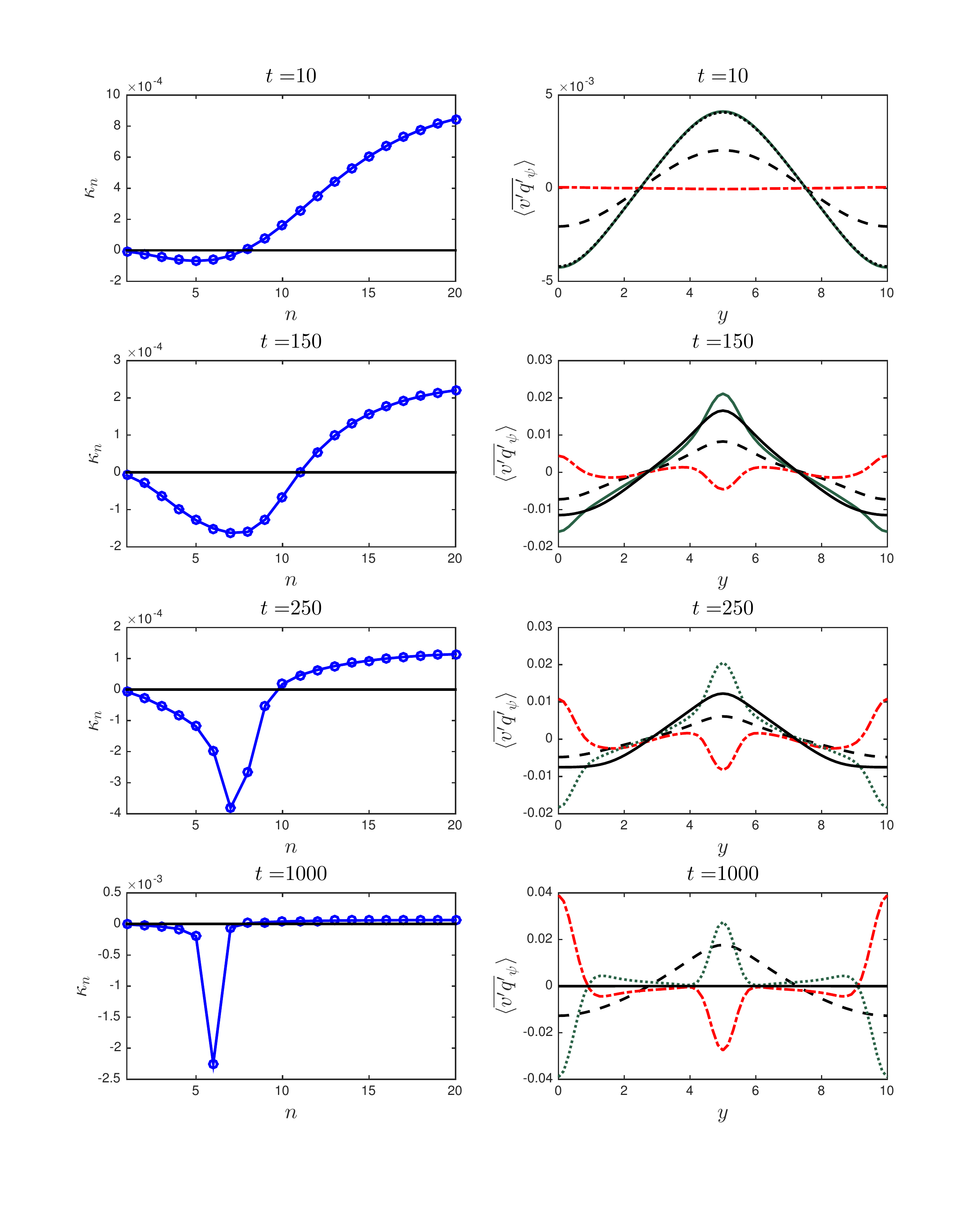}
        \caption{Left panels:  normalized rate of energy input to the jet   by each  wave $n$,
        $\kappa_n = \int_0^{L_y}U  \langle \overline{v' q'}_\psi\rangle_{n }\;\df y \big/ \int_0^{L_y}\tfrac1{2}U^2\;\df y$ ($\rm day^{-1}$),    for times $t=10,~150,~250,~1000$ as the S3T equilibrium
        is established starting from a small random initial jet
      structure. Right panels: the  latitudinal distribution of the  vorticity flux $\langle \overline{v' q'}_\psi \rangle$ (solid) (units: $11.57 ~\rm m\,s^{-1}\,day^{-1}$),
        the vorticity flux $\langle \overline{v' q'}_\psi\rangle_>= \sum_{n=12}^{56} \langle \overline{v' q'}_\psi\rangle_{n}$ from zonal
     waves $n\ge 12$ (dashed line), and  the vorticity flux $\langle \overline{v' q'}_\psi\rangle_< = \sum_{n=1}^{11}\langle \overline{v' q'_\psi}\rangle_{n}$
        from zonal waves  $n < 12$ (dotted line).   The structure  of
        the zonal velocity  at the corresponding time is indicated with a dashed line (amplitude arbitrary chosen to fit the graph).  Vorticity fluxes
        $\langle \overline{v' q' _\psi}\rangle_>$ are upgradient and are responsible for forming and maintaining  the jet,
        vorticity fluxes $\langle \overline{v' q'_\psi}\rangle_<$  are  downgradient opposing the jet  and these are responsible for the jet
        equilibration.   The S3T equilibrium is attained by $t=1000$ so that consistently $\langle\overline{v' q'}_\psi\rangle (y) = 0$.
       Energy loss from the jet at equilibrium  is concentrated  at $n=6$ and this energy loss is primarily balanced by energy input due to
      waves with $n\ge 8$.  Parameters: $L_y=10^4 ~\rm km$, $L_x = 8 \times 10^4 ~\rm km$, $\varepsilon=1$ corresponding to equal energy injection in both layers of
      $10^{-4}~ \rm W\,kg^{-1}$,   $r_m=0$ , $r_p=0.2~\rm day^{-1}$ and  $\tilde \beta= 6.9$.
}
\label{fig:vq}
\end{figure*}

The jet equilibrium is established by a robust feedback regulation arising from interaction between the mean equation and the perturbation covariance equation.
The mean jet, which is undamped, grows  from an arbitrarily small perturbation in the mean flow at first exponentially under the influence of the upgradient  fluxes induced by  shear straining of  the short waves.  This growth is progressively opposed and eventually
terminated by downgradient  fluxes associated with the arising of this
nearly neutral mode as the jet vorticity gradient sign change begins to be established (cf. Fig.~\ref{fig:vq}).  Extensive
experience with simulations has convinced us that this incipient instability closely constrains the jet amplitude to allow only relatively small vorticity gradient sign changes to occur.
It is widely recognized that the large vorticity gradient sign change in the NPJ observations poses a conundrum~\cite{Antunano-etal-2015}.
Within our model framework this discrepancy between the observed large vorticity gradient sign change in the observations of the upper layer of the NPJ  can only be resolved by regarding the observed jet equilibrium as an indirect observation of a larger effective value of $\beta$ in the unobserved lower layer of the  NPJ and in fact the S3T equilibrium jet with an appropriate choice of $\beta$, which is $\beta_{eff}=6.9 \beta_{sat}(74^\circ)$,
is  in close agreement  with the observed NPJ jet (cf. Fig.~\ref{fig:Unorm}b). We wish now to establish the dynamical argument compelling this conclusion.

\section{The equilibration mechanism underlying the robust scaling of  weakly damped turbulent jets}

We turn now to study in more detail the mechanism underlying the universal scaling of the structure and amplitude of  weakly damped turbulent
jets.

Note that in an undamped  jet  at equilibrium the perturbation momentum flux divergence  vanishes at each latitude,
$\overline {v'q'}(y) =0$ (cf. $t=1000$  right panel of  Fig.~\ref{fig:vq}).  Moreover, this requirement is independent of the stochastic excitation amplitude,
$\varepsilon$.  As previously mentioned,  for a given $\varepsilon$, perhaps from observational constraints,  the energy of the perturbation field at equilibrium  can  be shown to increase inversely  with  perturbation damping, $r_p$, to a good approximation.   These considerations imply invariant structure at equilibrium in the small $r_m$ limit for both the jet  (cf. Fig.~\ref{fig:Unorm}) and the perturbation turbulence component with only the  amplitude of the perturbation turbulence component varying and that variation being  as the ratio of excitation to damping, $\varepsilon /r_p$.

\begin{figure}
	\centering
       \includegraphics[width = \columnwidth,trim = 4mm 6mm 4mm 0mm, clip]{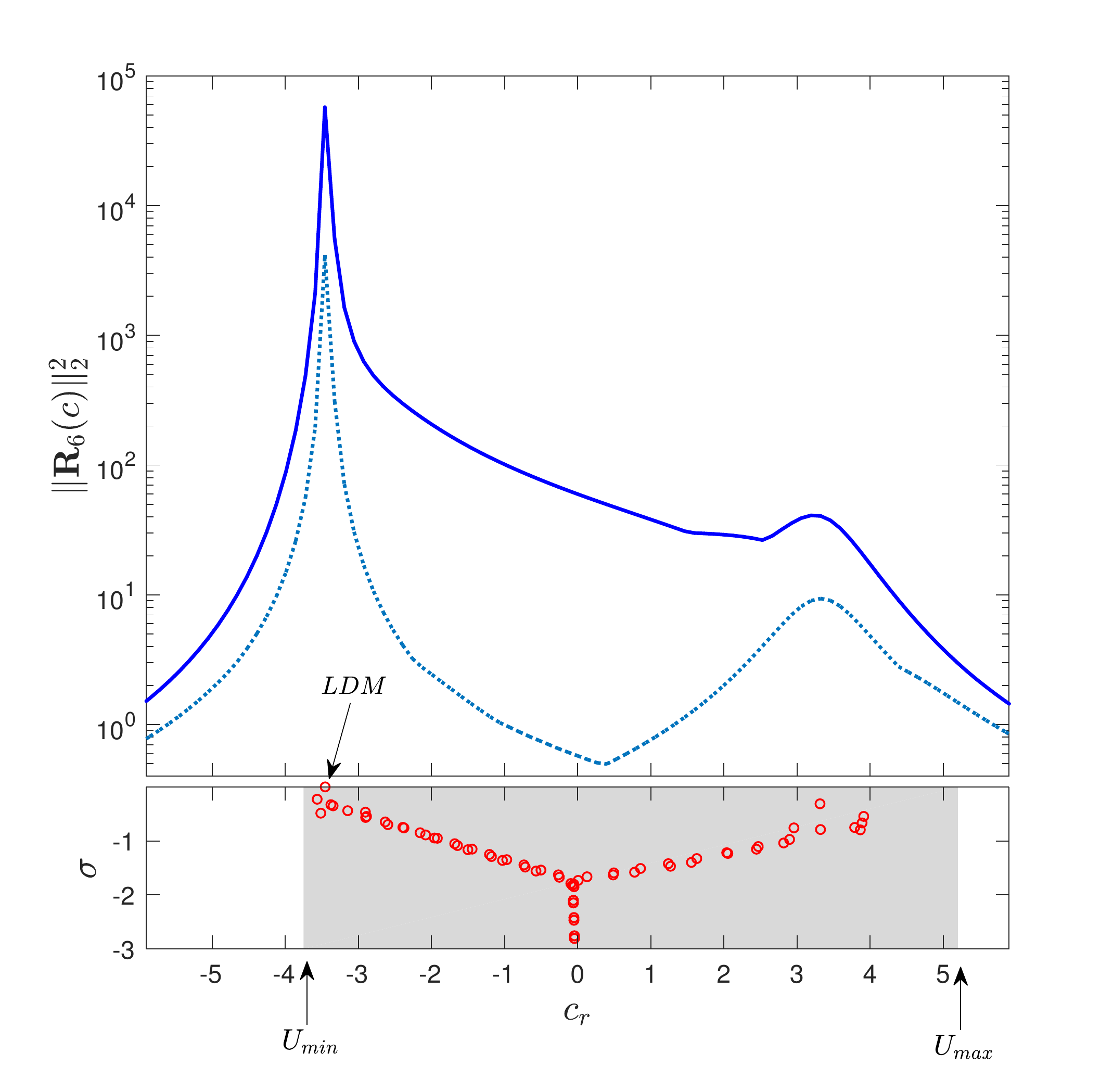}
        \caption{ Top panel: the square of the energy norm of the resolvent for $n=6$, $\| \R_6(c)\|_2^2$,  (solid)  as a function of phase speed, $c$,  indicating
         the maximum energy  amplification over all latitudinal structures $f(y)$ for unit energy harmonic forcing  of the form
         $f(y)\,e^{\ii k(x -c t)}$  with $k= 2 \pi n / L_x$. Also shown is the square of the energy norm of the equivalent normal
         resolvent which would obtain if the eigenfunctions of $\A_k$ were orthogonal (dotted). In the bottom panel is shown the
         growth  rate,  $\sigma$, and phase speed, $c_r$, of the modes of $\A_k$ for $n=6$ (circles). The dominant response
         arises in association with the nearly neutral  least damped mode (indicated LDM)  with $c=-3.44$
        and critical layer inside the retrograde jet.
 Although the secondary peak in the prograde jet associated with the mode with phase speed $c=3.32$
  arises in association with a prominent sign change in vorticity gradient the associated modes are not  significant in giving rise to the fluxes responsible for equilibrating the jet.
 The equivalent normal response is substantially smaller than the actual response indicating that non-normal interactions dominate the energetics at all phase speeds and most importantly at the phase speed  of the LDM where the maximum interaction occurs.  Quantities are non-dimensional and  parameters as in Fig.~\ref{fig:vq}.}
\label{fig:resolvent}
 \end{figure}

In addition to its anomalously large amplitude
given the small planetary value of $\beta$ available to stabilize it, the NPJ is also remarkable for supporting a prominent wavenumber six perturbation with  the amplitude required to distort the jet into a distinct hexagonal shape.  S3T equilibria comprise both the structure of the mean jet and the perturbation covariance from which information on perturbation structure can be determined.  A temporal sequence showing establishment of the equilibrium structure of the NPJ starting from a random initial condition and assuming the inferred $\tilde \beta= 6.9$ is shown in Fig.~\ref{fig:vq}.  As is generally found, the upgradient fluxes are produced by the short waves~\citep{Kitamura-Ishioka-2007,Farrell-Ioannou-2009-equatorial}, which in the case of the NPJ means waves with $n >12$.
The transfer of perturbation energy of wave $n$
to the mean is quantified
by
$\kappa_n = \int_0^{L_y}U  \langle \overline{v' q'}_\psi\rangle_{n }\;\df y \big/ \int_0^{L_y}\tfrac1{2}U^2\;\df y$ ($\rm day^{-1}$) and plotted in Fig.~\ref{fig:vq}.
Conversely, downgradient fluxes are produced by the long waves with $n \le 8$. The upgradient fluxes are associated with shear straining of the short waves which can be regarded as a
mechanism resulting in  a negative viscosity in that it produces upgradient momentum flux proportional to the velocity gradient~\citep{Bakas-Ioannou-2013-jas} as is observed in the atmospheres of
both Jupiter and Saturn~\citep{Ingersoll-etal-2004,Salyk-etal-2006, Delgenio-etal-2007}.  This shear straining mechanism accelerates both the prograde and retrograde jets. When the homogeneous turbulence is perturbed by a random mean  jet  these upgradient fluxes immediately cause the jet to grow in the form of  the most unstable S3T eigenmode (cf. time $t=10$ in Fig.~\ref{fig:vq})
\citep{Farrell-Ioannou-2007-structure,Farrell-Ioannou-2008-baroclinic}.  As the jet amplitude increases
the retrograde jet progressively exceeds the  speed of the slower retrograde Rossby modes and these sequentially obtain critical layers inside the retrograde jet~\citep{Kasahara-1980}. Approach to and attainment of a critical layer by these  modes is accompanied by increasing modal as well as non-normal energetic interaction with the jet resulting in increasing downgradient fluxes opposing the growth of the S3T jet eigenmode so as to eventually establish a nonlinear equilibrium. As neutral stability is approached with increasing jet amplitude these fluxes come into balance establishing by time $t=1000$ the stable fixed point equilibrium turbulent jet structure and associated perturbation covariance that together constitute a complete solution for the turbulent state at second order as shown in Fig.~\ref{fig:vq}. The equilibrium shown in Fig.~\ref{fig:vq}
reveals the mechanisms responsible for forcing
the large scale jet
and also the mechanism of dissipation at large scales that equilibrates the jet:
the jet receives energy from the small scale incoherent components of the turbulence, rises to finite amplitude and equilibrates by
transferring energy to the wave six  structure, which provides the sink for the jet energy.
Note that in this planetary scale turbulence regime
both the upscale energy transfer  forcing the jet and
the downscale energy transfer to the wave six mode
regulating its amplitude  are nonlocal in spectral space and neither involves a turbulent cascade.

\begin{figure}
\centering
\includegraphics[width = \columnwidth]{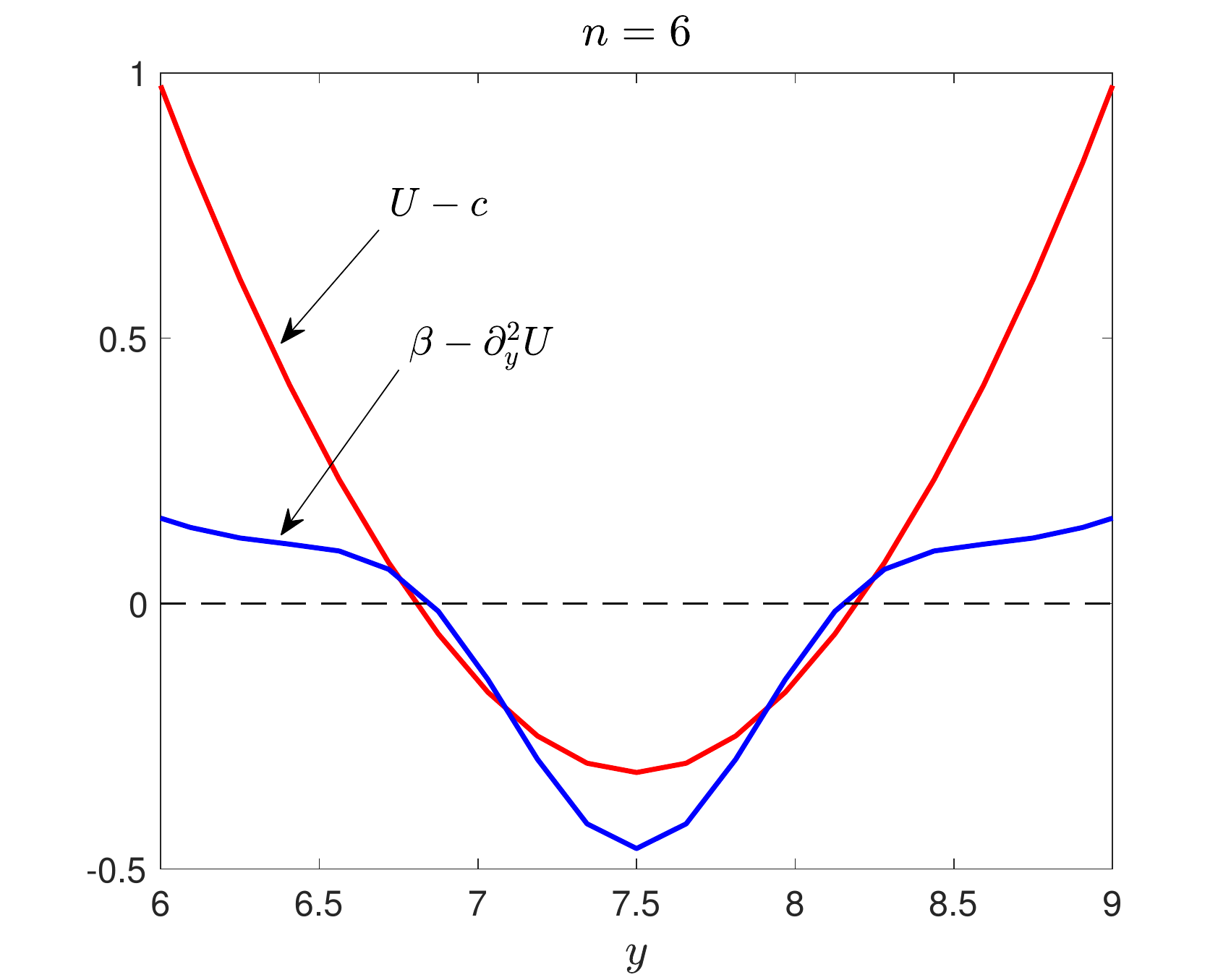}
\caption{Detail near the minimum velocity of the retrograde S3T  jet at equilibrium with $\tilde\beta=6.9$ showing the critical layer of the LDM, $U(y)-c_r$, where $c_r$ is the phase speed of the nearly neutral LDM of $\A_k$ for $n=6$. Also shown is the jet vorticity gradient, $\beta-\partial_y^2 U$, as a function of $y$. Note that the perturbation-mean interaction has stabilized the LDM by nearly eliminating the vorticity gradient in the vicinity of this mode's critical layer while leaving a small change in vorticity gradient sign between the mode's critical layers. Such changes in sign of the vorticity gradient are a commonly observed feature of turbulent jets in planetary atmospheres. Quantities are non-dimensional and parameters as in Fig.~\ref{fig:vq}.}\label{fig:Qy6}
\end{figure}

Further insight into the dynamics can be obtained by calculating the eigenvalues and the resolvent of the perturbation dynamics operator $\A_k({\U})$ as the jet structure $\U$ evolves toward equilibrium.   The eigenvalues reveal the approach of the retrograde modes' phase speeds to the speed of the retrograde jet and the associated  decrease in mode damping rate which is indicative of energetic
interaction with the jet and diagnostic of downgradient momentum flux by the mode.
The resolvent of $\A_k({\U})$ provides more information by revealing the response of the dynamics to the turbulent excitation at each mode phase speed, $c$. We use the energy norm to measure the response of the dynamics and the non-normality of the dynamics are defined with respect the energy norm  (cf.~Ref.~\cite{Farrell-Ioannou-1996a}). The resolvent  of perturbations with
 zonal wavenumber
$k = 2 \pi n / L_x$  perturbations is :
\begin{equation}
\R_{n}(c) = -( \ii k  c ~\I + \A_{k})^{-1}~.
\end{equation}
 The square norm of the resolvent of $\A_k({\U})$, which is the energy spectrum as a function of phase speed
for  temporally and spatially delta correlated excitation,  is shown together with the spectrum of the modes of
$\A_k({\U})$ for the equilibrium jet structure, $\U$, in   Fig.~\ref{fig:resolvent}.
There is a dominant nearly neutral Rossby mode with wavenumber six that is responsible for most of the downgradient momentum flux balancing the upgradient shear straining fluxes from the short waves (cf.~Fig.~\ref{fig:vq} at $t=1000$). This wave dominates the response of the dynamics to perturbation when the jet is equilibrated as can be inferred from the resolvent. This dominance of wavenumber six in the perturbation variance  extends over a wide range in $\beta$ and therefore in equilibrium jet amplitude as shown in Fig.~\ref{fig:DeltaU}. Note that the phase speed of this mode has been incorporated into the retrograde jet  but that this mode remains stable  consistent with finiteness of the perturbation variance. The strong $\overline{v' q'}$ fluxes  associated with the maintenance of this mode have forced the gradient of the mean vorticity in the vicinity of its critical layer nearly to zero as shown in Fig.~\ref{fig:Qy6}. This nonlinear interaction provides an example of the mechanisms at play in the complex feedback stabilization process operating between the first and second cumulants in the S3T dynamics that results in  establishment of the equilibrium statistical state.  It is useful to regard this interaction as a nonlinear regulator that continuously adjusts the mean flow to a state that is in neutral equilibrium with the perturbation dynamics by enforcing vanishing of the mean vorticity gradient at the critical layers of the dominant perturbation modes in the retrograde jet.
This consideration explains why strongly excited jet equilibria in planetary atmospheres commonly exhibit easily observable changes in sign of the mean vorticity gradient without incurring instability: shear straining of the small turbulence components drives both the prograde and retrograde jets strongly producing the sign change while the regulator need only equilibrate any incipient modal instability by enforcing vanishing of the gradient at the mode critical layer while leaving a substantial vorticity sign change between the critical layers in the jet profile.  It is important to appreciate the crucial role of active  feedback regulation continually operating between the first and second cumulants in the SSD in maintaining the stability of this turbulent equilibrium jet-wave-turbulence state.
If one were simply to postulate a jet profile very much like the observed it would be extremely unlikely to have by chance the vanishing gradient of vorticity precisely at the critical layers of the mode required for stability of the jet structure.  Viewed another way, the existence of strong jet equilibria with the structure seen in planetary atmospheres requires that an active feedback regulation be operating to maintain their stability.  Note that the mechanism of vorticity mixing  in the retrograde jets could not result in a negative vorticity maximum as is seen in both observations and our simulations.

While it is tempting to regard the downgradient momentum fluxes arising from the dominant mode itself as being primarily responsible for opposing the upgradient fluxes by the small waves, the resolvent tells a different story.  Shown in Fig.~\ref{fig:resolvent} is both the response of the jet dynamics to perturbation as a function of phase speed and the response that would be produced by the modes assuming they were independent; that is, assuming the modes to be orthogonal in energy.  This so called equivalent normal response reveals that the energetics and therefore the fluxes associated with the LDM at $n=6$ are being produced by non-normal interaction among the modes rather than by the $n=6$  mode by itself.
In fact it is generally the case in systems non-normal in energy
that  transient growth of the adjoint of a mode is responsible for establishing  a mode's equilibrium amplitude when it is excited stochastically rather than growth of the mode itself~\citep{Farrell-Ioannou-1996a}.

 \begin{figure}
	\centering
       \includegraphics[width = \columnwidth,trim = 12mm 0mm 12mm 5mm, clip]{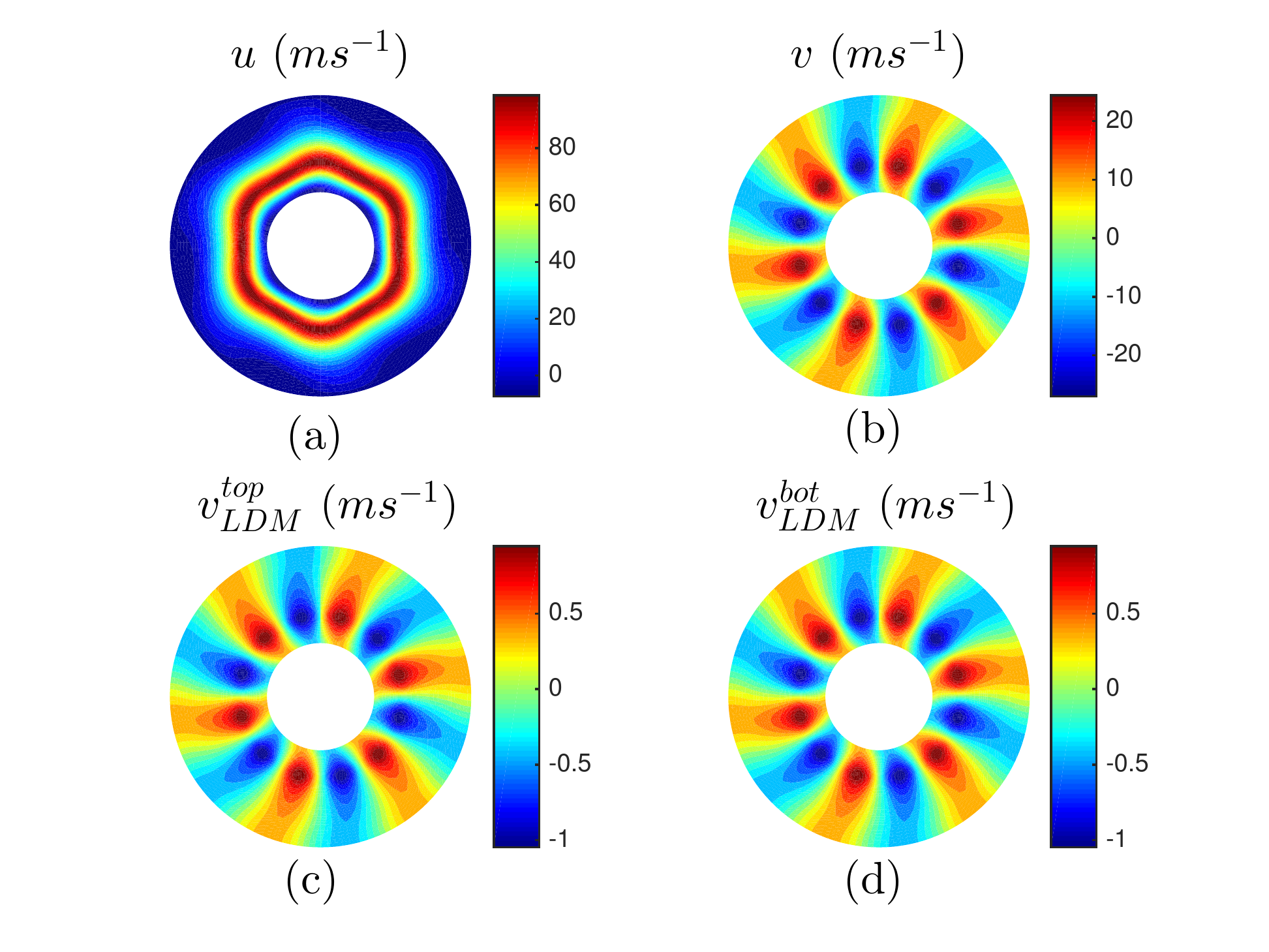}
        \caption{ The predicted  NPJ velocity structure in a polar rendering of the S3T model channel (the external circle
         of the annulus  corresponds to latitude $70^o$ and the inner circle to  $82^o$).
        Panel (a): contours of the total zonal velocity obtained by adding the S3T equilibrium zonal mean velocity
        and the zonal velocity of the first POD mode obtained from eigenanalysis of the equilibrium perturbation covariance at zonal wavenumber $n=6$.  This mode accounts for
         $99.5\%$
        of the perturbation  energy at this wavenumber.   The amplitude of the wave is obtained from the associated eigenvalue of the equilibrium perturbation covariance.
        The jet is barotropic as is the $n=6$ wave. Panel (b):  contours of the meridional velocity of the first POD of the perturbation covariance at zonal
        wavenumber $n=6$.
        Panels (c) and (d) show respectively the structure in
         the top and bottom layer of the least damped mode (LDM) of  $\A_k$ at $n=6$ (cf. Fig.~\ref{fig:resolvent}).
        It is clear from this comparison that
        the LDM is barotropic and the  POD has the structure of this mode. Parameters are as in Fig.~\ref{fig:vq}.}
\label{fig:vpolar}
 \end{figure}


  \begin{figure*}
	\centering
       \includegraphics[width = .5\textwidth]{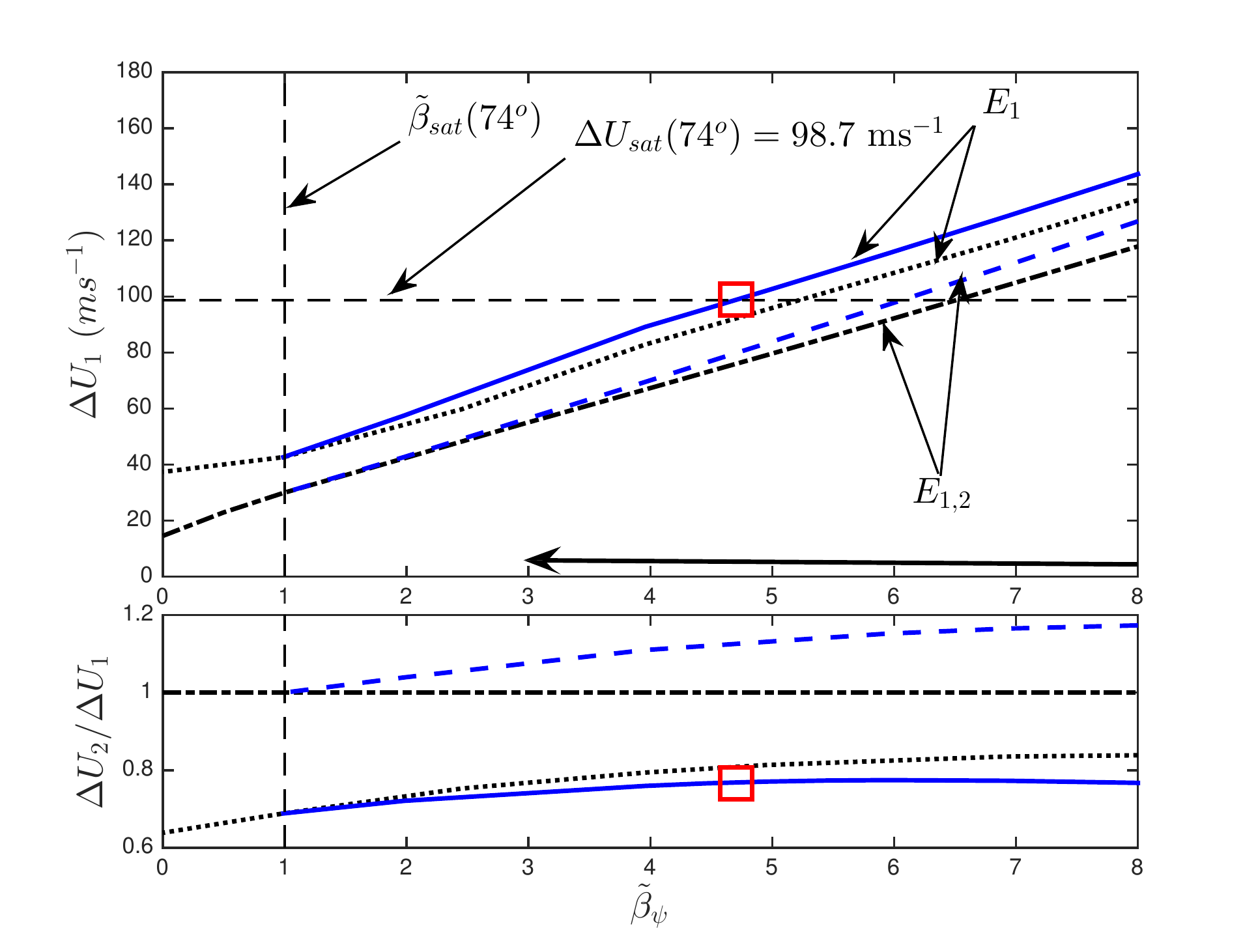}
        \caption{ Top panel: S3T equilibrium jet amplitude, $\Delta U$, as a function of  the normalized
        barotropic $\beta$ component $\tilde \beta_\psi = \beta_\psi/ \beta_{sat}(74^\circ)$ for Saturn's  NPJ ($L_y=10^4~\rm km$).
         $\Delta U_1$ is the difference between the maximum and the minimum jet velocities of the top layer. Lines  $E_{1,2}$ indicate equilibria obtained when both layers are equally excited, lines $E_1$ indicate equilibria obtained
          when only the top layer is excited.  The solid and dashed  lines  are equilibria with bottom layer $\beta_2=2 \beta_\psi - \beta_{sat}$  and top layer
          $\beta_1=\beta_{sat}$. The dash-dot and dotted lines indicate equilibria in which both layers have the same $\beta_\psi$, as in Fig.~\ref{fig:DeltaU}.  Also indicated are the
          planetary values of $\tilde \beta_{sat}(74^\circ)$ and  the
          observed $\Delta U_{sat}(74^\circ) = 98.7~\rm m\,s^{-1}$ from Ref.~\citep{Antunano-etal-2015};  boxes indicate the point consistent with the NPJ observations . The arrow indicates the range of $\beta_\psi$
          for which the perturbation energy is concentrated in a single  nearly neutral wave
        with phase speed near that of the jet minimum, which for the NPJ channel corresponds to $n=6$.
         Scaling of  $\Delta U$ with $\beta L_y^2$ is  highly accurate in this region.
        Parameters: $r_m=0 $, $\varepsilon=1~\rm W\,kg^{-1}$, $L_R = 10^3~\rm km$, $r_p=0.2~\rm day^{-1}$.
       Bottom panel: The resulting baroclinicity of the equilibria, measured as the ratio of $\Delta U_2/\Delta U_1$ in each layer.
       Barotropic flows, such as occur when the excitation and $\beta$ are the same in both layers (cases  $E_{1,2}$) have $\Delta U_2/\Delta U_1=1$ (dash-dot line).
       In all other cases the equilibria are slightly baroclinic. When the  upper layer alone is excited ($E_1$) the upper layer jet is stronger than the jet in the bottom layer (solid and dotted lines). If both layers are excited and the value of $\beta$ in the bottom layer is greater than that
       in the top layer, the bottom jet is the stronger (dashed line).}
\label{fig:DeltaUbca}
 \end{figure*}

Because of its dominance the structure of the $n=6$ perturbation can be obtained as the first eigenmode of the perturbation covariance (referred to variously as the leading proper orthogonal decomposition (POD)  or empirical orthogonal function (EOF) mode).  This structure is shown in Fig.~\ref{fig:vpolar}b and its superposition on the jet with  amplitude obtained as the RMS of its variance as obtained from the associated POD eigenmode is shown in Fig.~\ref{fig:vpolar}a. Consistent with the resolvent diagnostic discussed above, this mode has nearly the same barotropic structure as the least damped mode of the linear perturbation dynamics,
shown in Fig.~\ref{fig:vpolar}c,d.   The prediction of the theory
that the wave six phase speed  be just inside
the retrograde jet is in agreement with observations \citep{Antunano-etal-2015}.

 The amplitude of the wave six mode is proportional to
$\varepsilon /r_p$ and we have chosen physically plaussible values for these unknown parameters,
$\varepsilon=1$ and $r_p=0.2~\rm day^{-1}$, corresponding to excitation of
$10^{-4}~\rm W\,kg^{-1}$. However, other combinations of
$\varepsilon$ and $r_p$ with
$\varepsilon/r_p=5$ lead to equilibria very close to those obtained with
these  parameter values.

\section{The deep stable layer NPJ model}

Our study of the dynamical consequences for jet formation and equilibration of varying the planetary value of $\beta$  reported above has the advantage of allowing the underlying  mechanisms and the associated scaling to be understood in a simple context.  However, the physical mechanism by which an   effective value of $\beta$ differing from the planetary value enters the dynamics of the NPJ  is likely to be a poleward sloping  surface of concentrated downward increase in static stability  underlying the deep jet   inducing
the dynamic analogue of a topographic  $\beta$ effect in the lower layer. This equivalent sloping lower boundary could be  associated with constitutive, convective and/or dynamical processes analogous to those which are responsible for maintaining the  Earth's tropopause but lacking  observation it is not possible to identify the specific processes responsible.

A topographic $\beta$ effect results in different values of effective $\beta$ in the two layers rather than the same value in both as was appropriate when varying the planetary value of $\beta$  in the theoretical development above. In the top layer the zonal mean potential vorticity
gradient (PV gradient) is $Q_{1y}= \beta_{sat} - \partial_y^2 U_1 + \lambda^2 (U_1-U_2)$, with $\beta_{sat}$  designated the planetary value of $\beta$, and in the bottom layer it is  $Q_{2y}= \beta_{sat} +\beta_h- \partial_y^2 U_2 - \lambda^2 (U_1-U_2)$, in which the planetary value of $\beta$ has been designated $\beta_{sat}$ and the topographic value $\beta_{h}$. The equations~\eqref{eq:BRCL} are modified as follows:

 \begin{figure}
	\centering
       \includegraphics[width = \columnwidth]{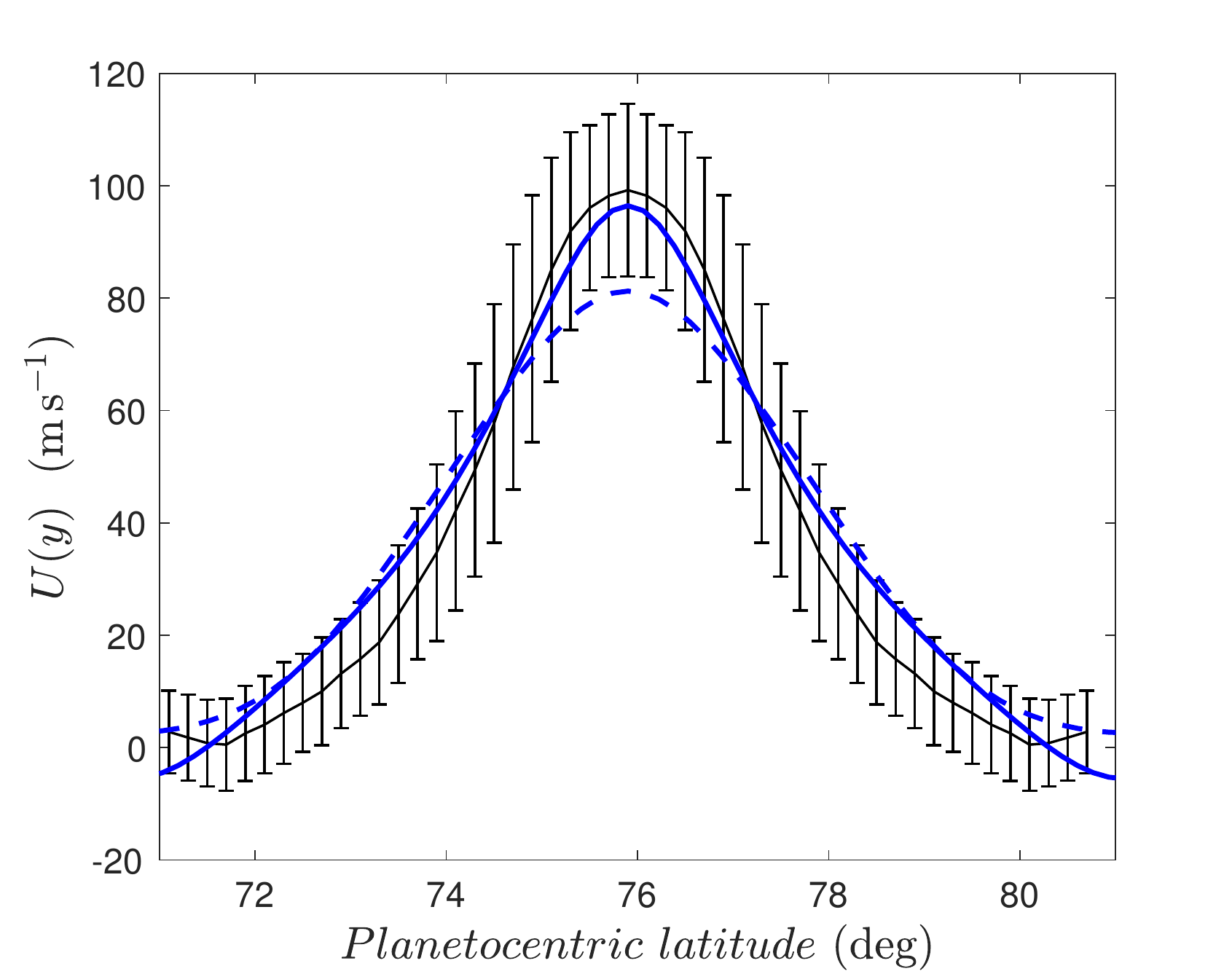}
        \caption{For the deep stable layer NPJ model: equilibrium jet in the top layer (thick solid line) and bottom layer (thick dashed line).
        Also shown is the symmetrized observed NPJ from~Ref.~\citep{Antunano-etal-2015} (thin dashed line with one standard deviation error bounds).
        The S3T equilibria are obtained with stochastic forcing of the top layer, $r_m=0$, $\lambda=1$ and $\tilde \beta_\psi=4.9$ corresponding to topographic component
        $\tilde \beta_{h} = 8.8$ in the bottom layer. This equilibrium profile is insensitive to variations in $\varepsilon$, $r_p$ and $\lambda$.}
\label{fig:Ub1b2f5}
 \end{figure}

\begin{subequations}
\label{eq:BRCL12}
\begin{align}
\label{eq:BRCL12a}\partial_t \Del \psi + &J(\psi,\Del\psi) + J(\theta,\Del\theta)  \nonumber \\
&+  \partial_x\beta_\psi\psi+\beta_\theta \partial_x\theta = -r \Del \psi +\sqrt{\varepsilon}\,f_\psi~,\\
\partial_t \Del_\la \theta + &J(\psi,\Del_\lambda\theta) + J(\theta,\Del\psi)  \nonumber \\
&+ \beta_\psi \partial_x\theta +\beta_\theta \partial_x\psi= 
-r \Del_\la \theta+\sqrt{\varepsilon}\,f_\theta~,
\end{align}\end{subequations}
with the barotropic, $\beta_\psi$, and baroclinic, $\beta_\theta$, defined as:
\begin{equation}
\beta_{\psi} \equiv \beta_{sat} + \beta_h/2~,~~\beta_{\theta} \equiv \beta_{sat} - \beta_h/2~.
\end{equation}

The S3T system is modified accordingly.
From ~\eqref{eq:BRCL12a}  we see that  the effective value of $\beta$ seen by waves with primarily barotropic structure such as  the planetary wave $n=6$ that is implicated in the dynamics of the NPJ equilibration is the sum of the planetary value  $\beta_{sat}$ and half the topographic value, $\beta_h$.
Results similar to those shown in Fig.~\ref{fig:DeltaU} for the case of equal $\beta$ in both layers are shown in Fig.~\ref{fig:DeltaUbca} for the case of a topographic $\beta$ effect
with
$\beta=\beta_{sat}+\beta_{h}$ in the bottom layer and  $\beta=\beta_{sat}$ in the top layer. Even with equal excitation of the background turbulence in both layers
 ($E_{12}$ case)  both the jet and the waves become slightly
baroclinic as seen in Fig.~\ref{fig:DeltaUbca} and the jet  in this case obtains a higher equilibrium amplitude in the lower layer consistent
with the higher effective value of $\beta$ there.
However, when excitation is restricted to the top layer ($E_1$ case in Fig.~\ref{fig:DeltaUbca}) the jet in the top layer is the stronger.

 \begin{figure}
	\centering
       \includegraphics[width =\columnwidth,trim = 12mm 0mm 12mm 5mm, clip]{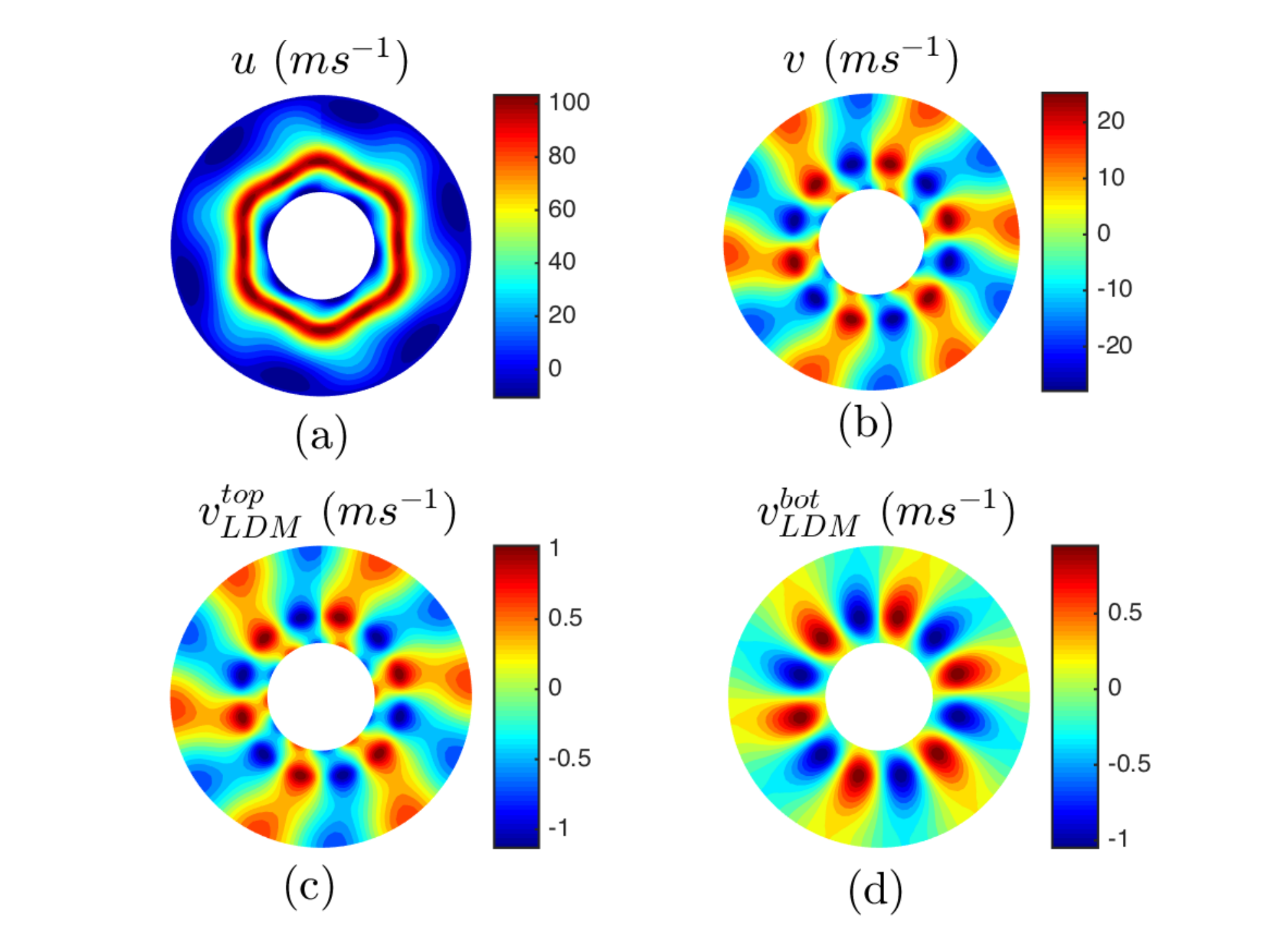}
        \caption{For the deep stable layer NPJ model  shown in Fig.~\ref{fig:Ub1b2f5}:  NPJ structure   in a polar rendering of the
        channel (the external circle
         of the annulus  corresponds to latitude $70^o$ and the inner circle to  $82^o$).
        Panel (a): contours of the total zonal velocity as obtained by adding to the S3T equilibrium zonal mean velocity
        the zonal velocity of the first POD of the equilibrium perturbation covariance at zonal wavenumber $n=6$ which accounts for $99.7\%$ of the perturbation energy
        at this wavenumber. The amplitude of the wave is obtained from the associated eigenvalue of the equilibrium covariance. The jet and the $n=6$ wave are slightly baroclinic. Panel (b):
        contours of the meridional velocity of the first POD of the perturbation covariance at zonal wavenumber $n=6$.
        Panels (c) and (d) show respectively the structure of the least damped mode (LDM) of $\A_k$ for $n=6$ (indicated in Fig.~\ref{fig:resolvent5}) in the top and bottom layer
        (left and right panels respectively).
       The POD has the structure of this mode and both are  slightly baroclinic. For $\varepsilon=0.7$, other  parameters as in Fig.~\ref{fig:Ub1b2f5}.}
\label{fig:vpolar5}
 \end{figure}

  \begin{figure}
	\centering
       \includegraphics[width =\columnwidth,trim = 4mm 6mm 4mm 0mm, clip]{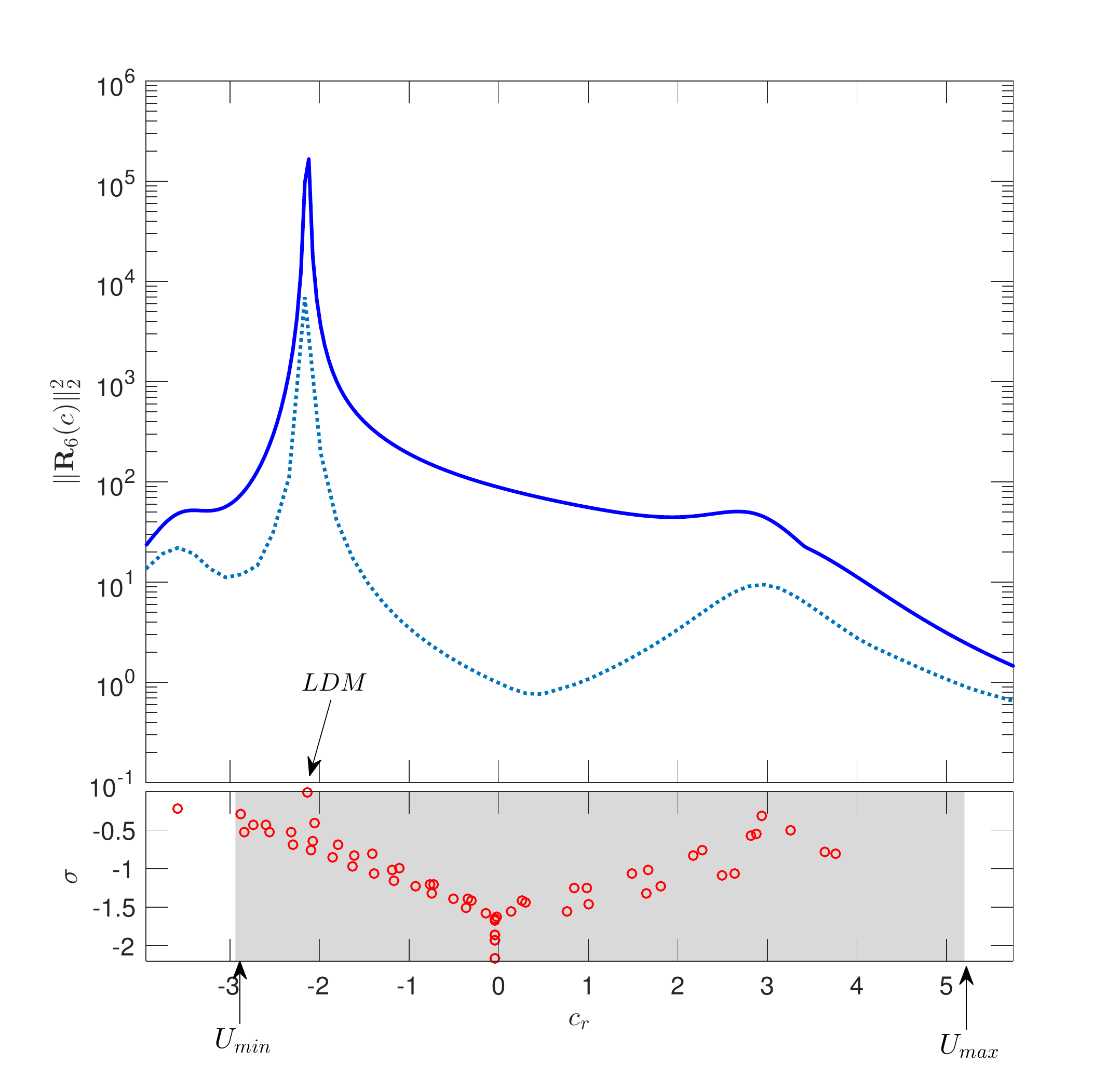}
        \caption{For the deep sloping stable layer NPJ model equilibrium shown in Fig.~\ref{fig:Ub1b2f5}:  in the top panel is shown the square of the energy norm of the resolvent for wave $n=6$, $\|\R_6(c)\|_2^2$ (solid), together with the equivalent normal response (dotted) as a function of phase  speed, $c$. The dominant response arises from the nearly neutral least damped mode (indicated LDM) with $c=-2.14$ and critical layer inside  the retrograde jet. Note the secondary peak caused by  the  non-normal excitation of the mode with $c=2.93$ and critical layer inside the prograde jet (see  Fig.~\ref{fig:ucritical}). In the bottom panel is shown the growth rate, $\sigma$, and phase speed, $c$, of the modes of $\A_k$ for $n=6$ (circles). The equivalent normal response is substantially smaller than the actual response indicating that non-normal interactions dominate the energetics at all phase speeds and most importantly at the phase speed  of the LDM where the maximum interaction occurs. Quantities are non-dimensional and  parameters as in Fig.~\ref{fig:Ub1b2f5}.
}
\label{fig:resolvent5}
  \end{figure}

  \begin{figure}
	\centering
       \includegraphics[width = .5\textwidth]{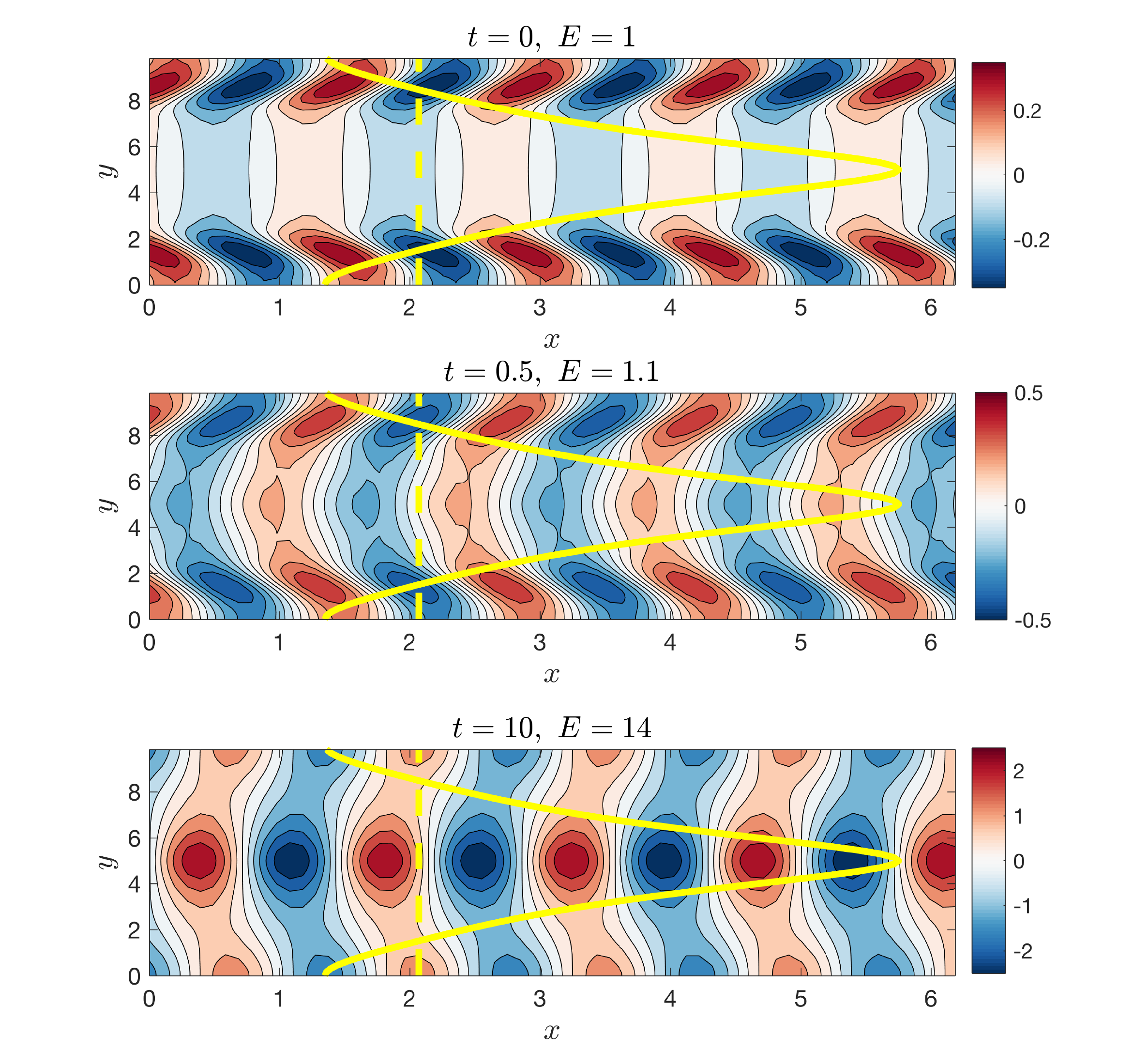}
        \caption{For the deep sloping stable layer NPJ model equilibrium shown in Fig.~\ref{fig:Ub1b2f5}: establishment of the $n=6$ LDM by its optimal excitation in energy structure.
        Shown is the evolution of the optimal as indicated in  the top layer streamfunction. Upper panel: the optimal perturbation at $t=0$ with energy density $E=1$. Middle panel:~the  evolved optimal at $t=0.5$. Bottom panel: the evolved optimal at $t=10$ at which time it has assumed the structure of the LDM with energy $E=14$. Contours indicate  non-dimensional values of the streamfunction. Also shown is the equilibrium jet (solid), which has been scaled to fit, and the phase speed of the LDM (dashed). While the amplitude of the LDM is concentrated in the prograde jet, its  optimal excitation  is concentrated near  its critical layer in  the retrograde jet. The non-normality of the dynamics is indicated by both the large excitation  of this stable structure  and the substantial structural change during the evolution of the optimal. }
\label{fig:adj}
\end{figure}
\begin{figure*}
	\centering
       \includegraphics[width = .6\textwidth,trim = 0mm 6mm 0mm 0mm, clip]{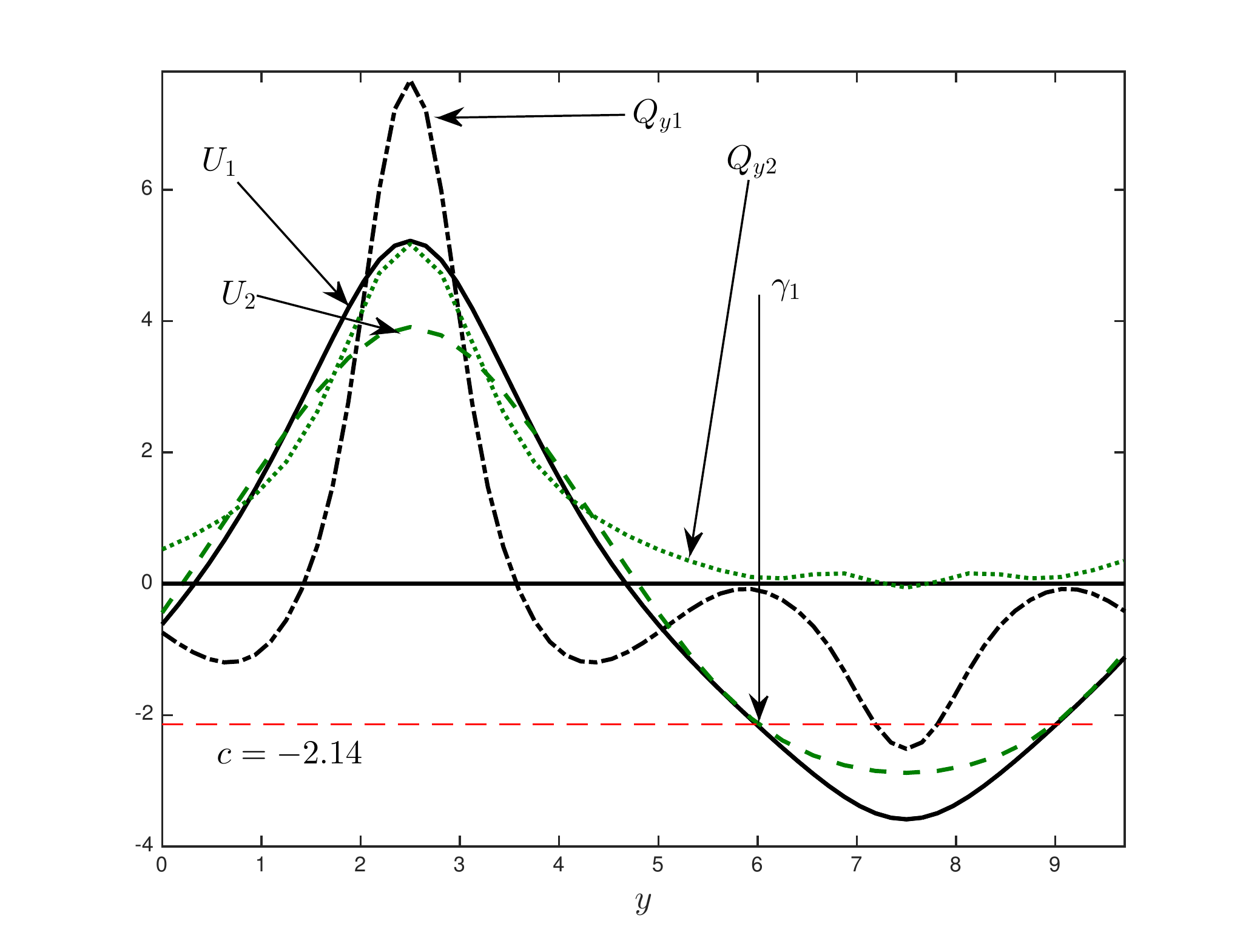}
        \caption{For the deep sloping stable layer NPJ S3T equilibrium shown in Fig.~\ref{fig:Ub1b2f5}: top layer velocity, $U_1$ (solid), bottom layer velocity, $U_2$ (dashed), corresponding top layer
        PV gradient, $Q_{y1}$ (dash dot)  and bottom layer PV gradient, $Q_{y2}$ (dot).
        Also indicated is the phase velocity of the LDM  with wavenumber $n=6$ that is the dominant perturbation in producing downgradient fluxes.
        Note that this LDM has eliminated the PV gradients  in the vicinity of its critical layers in the top and bottom  layers indicated by $\gamma_1$.  Elimination of the PV gradient at the critical layer is responsible for suppressing  instability of this LDM. The strong reversal in $Q_{y1}$ at the center of the retrograde jet resulting from the upgradient vorticity fluxes produced by the  $n>8$ waves remains as this region of PV gradient reversal is not opposed
         by the LDM which has been stabilized by elimination of the PV gradient at its critical layers.  Such reversals are commonly observed in planetary atmospheres.}
\label{fig:ucritical}
%
%
%
	\centering
       \includegraphics[width = .6\textwidth]{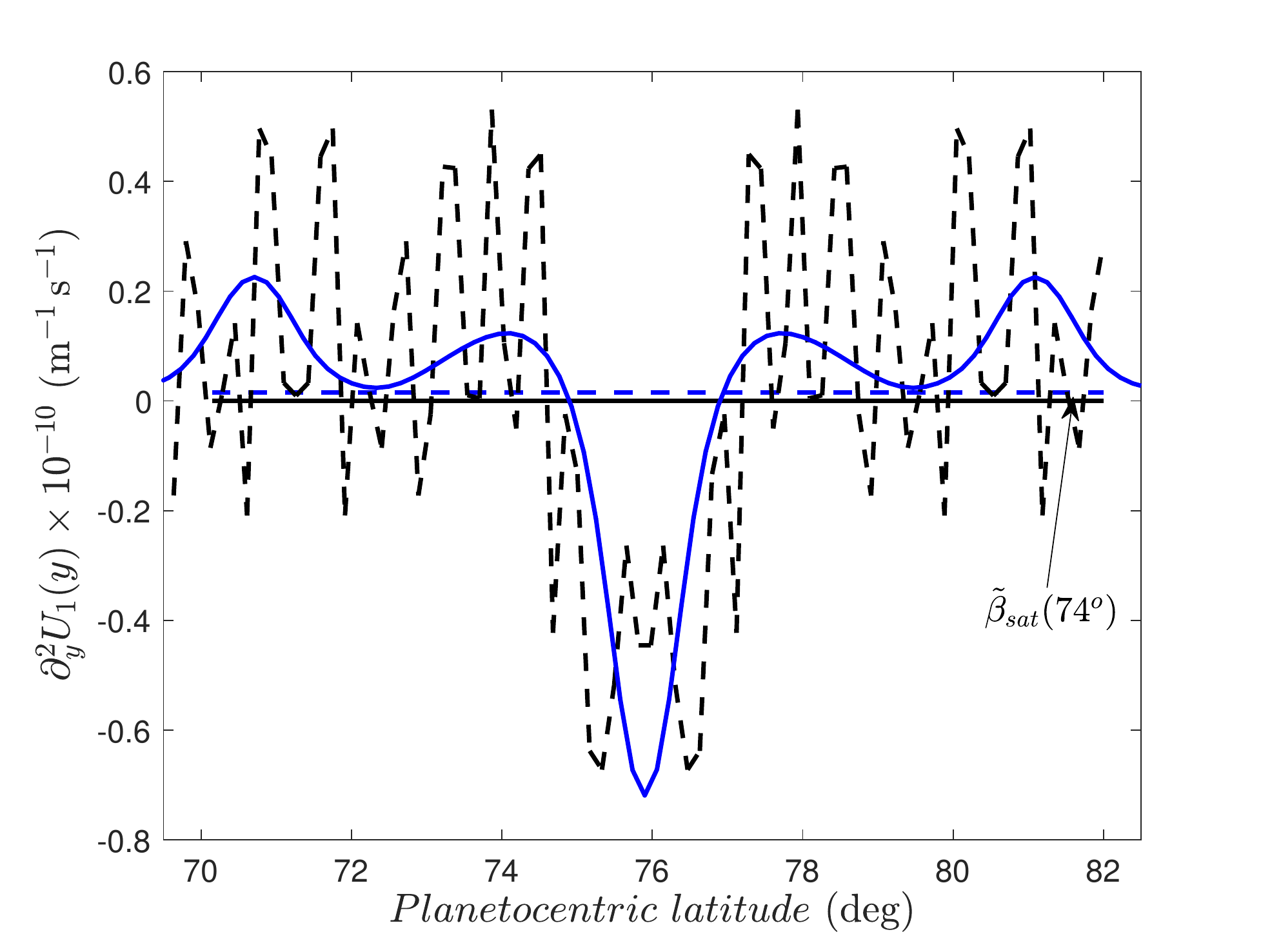}
        \caption{The dimensional curvature of the predicted  NPJ velocity structure   of the top layer jet of the S3T equilibrium shown in Fig.~\ref{fig:Ub1b2f5} (solid) plotted against
        the curvature obtained from  the raw data of Ref.~\citep{Antunano-etal-2015} (dashed line).
        Also indicated is the relatively insignificant planetary $\beta_{sat}(74^\circ)$. Although noisy
        the data suggests both the
        curvature reversal in the vicinity of the prograde jet, which the theory predicts to be related to the subdominant prograde mode with
        non-dimensional $c=2.93$ that is responsible for the secondary maximum in the resolvent response in Fig.~\ref{fig:resolvent5},  and also the   dynamically significant second maximum of the curvature in the retrograde section of  the jet (at $71^o$ N)
        predicted by the theory to lie between  the critical layers of the dominant $n=6$ wave  with non-dimensional phase speed c= -2.14.   }
\label{fig:D2Uobs}
 \end{figure*}

Assuming only the top layer is excited a topographic $\beta$ component in the lower layer of $\beta_{h} = 8.8 \beta_{sat} = 1.44 \times 10^{-11} ~\rm m^{-1}\,s^{-1}$ when combined with the  planetary $\beta_{sat}(74^\circ)$ results in consistency with the NPJ  upper layer observations as indicated in Fig.~\ref{fig:DeltaUbca}.
With a layer depth equal to  one scale height on Saturn, $H=42.1 ~\rm km$, the required lower layer slope is
\begin{eqnarray*}
{h_y}&=&\frac{\beta_{h} H }{f}=\frac{(14.4 \times10^{-12}~\rm m^{-1} s^{-1}) \times ( 42.1 \times 10^3~\rm m)}{3.1\times 10^{-4}~\rm s^{-1} }\\
&=&2 \times 10^{-3}~,
\end{eqnarray*}
implying a  density surface dynamically equivalent to the bottom boundary sloping   downward toward the pole over the channel width of
$10\,000 ~\rm km$ from  $74.5~\rm km$
to $93.9~\rm km$ (measured from the top boundary). 
We remark that   the prominent $24^\circ\;\rm N$ jet of Jupiter conforms with the lightly damped jet scaling using the planetary value of $\beta_{jup}(24^\circ)$, as shown
 in Fig.~\ref{fig:Unorm}  which result verifies previous findings~\citep{Farrell-Ioannou-2008-baroclinic}.
This suggests that the stable lower layer inferred to underly Saturn's NPJ is
peculiar to the anomalous thermal and dynamical conditions observed near Saturn's  Pole.

The S3T jet structure in the upper layer with the parameter values given above, which were chosen to model  the NPJ,  is shown in Fig.~\ref{fig:Ub1b2f5}.
 A polar representation of the jet and the  associated $n=6$ wave is shown in Fig.~\ref{fig:vpolar5}.
The jet equilibration process is essentially similar to that described in the previous section in which the planetary value of $\beta$ was varied.  As in the previous case, most of the perturbation variance is concentrated in the nearly neutral wave six which has a critical layer inside the retrograde jet as can be seen in the resolvent response of the $n=6$ wave shown in Fig.~\ref{fig:resolvent5}. Again, this nearly neutral wave with critical layer inside the retrograde jet and with primarily non-normal energetics
is in large part  responsible for producing the upgradient flux to exactly cancel the downgradient flux produced by the
waves $n \ge 8$ resulting in establishment of the turbulent equilibrium.
 In Fig.~\ref{fig:adj} is shown three snapshots  of  the temporal development
of the optimal initial condition for exciting the LDM demonstrating the dominance of non-normal dynamics in the establishment of this mode. It is interesting  to note that the optimal excitation  of this mode, which is its adjoint~\citep{Farrell-1988a,Farrell-Ioannou-1996a}, is initially concentrated in the retrograde  jet (cf. Fig.~\ref{fig:adj}). As in the previous example and  as is required for neutrality, the PV gradient in both the upper and lower critical layers ($Q_{y1}$ and $Q_{y2}$) is eliminated, primarily by fluxes arising from the non-normal dynamics, as shown in Fig.~\ref{fig:ucritical}.

The potential vorticity gradient reversals
seen in Fig.~\ref{fig:ucritical}, which are observed in both Jupiter and Saturn and predicted by the S3T equilibria, provide a probe
of the dynamical processes at work in the equilibration of  these large amplitude jets.
As previously discussed in connection with the planetary $\beta$ variation example,   a strong reversal of $Q_{y1}$ near the minimum of the retrograde jet results from upgradient vorticity fluxes
produced by the  $n>8$ waves. The resulting region of potential vorticity gradient reversal  is not
opposed by the LDM which enforces its own stabilization by eliminating the PV gradient in the vicinity
of its critical layers leaving an extensive
region between these critical layers in which the potential vorticity gradient  is negative,  making the
flow over the whole domain violate the Rayleigh--Kuo necessary condition for instability (the relevant
criterion for the two-layer model is
the Charney-Stern criterion, but this reduces to the Rayleigh--Kuo criterion for nearly barotropic flows).
It should be noted that violation of the Rayleigh--Kuo criterion does not guarantee the instability
of the flow and so on a logical level the violation of the Rayleigh--Kuo criterion  observed in most  jets on Jupiter
and Saturn  does not present a conundrum, except for the fact that  in almost
every case in which the Rayleigh--Kuo condition is violated
the flow is found to be unstable. It is demonstrated in this paper that the process of adjustment to statistical equilibrium
actively modifies the equilibrium jets that violate the Rayleigh--Kuo condition to neutrality by placing the critical layer
of the incipient unstable wave in coincidence with the location at which the mean  potential vorticity vanishes,
in this way equilibrating the
instability.

%
%
Note that the bottom layer
PV gradient has also been eliminted in the region between  the critical layers of the $n=6$ wave.  However, in this case with excitation limited to the top layer the shorter waves  do not penetrate adequately into the bottom layer to produce sufficiently strong upgradient fluxes to form a PV gradient reversal.

Reversals in PV gradient in the retrograde jet are associated with positive curvature of the mean flow. The predicted curvature of the mean flow in the top layer is plotted in Fig.~\ref{fig:D2Uobs}  against raw observations from Ref.~\citep{Antunano-etal-2015}.
Discussion tends to center  on the maximum at the wings of the prograde
jet which is responsible for supporting modes associated with  the secondary maximum in the resolvent at phase speeds
inside the prograde jet seen in Fig.~\ref{fig:resolvent5}.
However, the modes associated with this prograde jet PV reversal are insignificant in the dynamics. The dynamically significant PV gradient reversal is that  between the critical layers of the dominant $n=6$ wave.
Neither of these features is well resolved in the data but we believe that better observations of the PV in the retrograde jets would resolve the dynamically significant reversal between the $n=6$ critical layers.


\begin{figure}
	\centering
       \includegraphics[width = .5\textwidth]{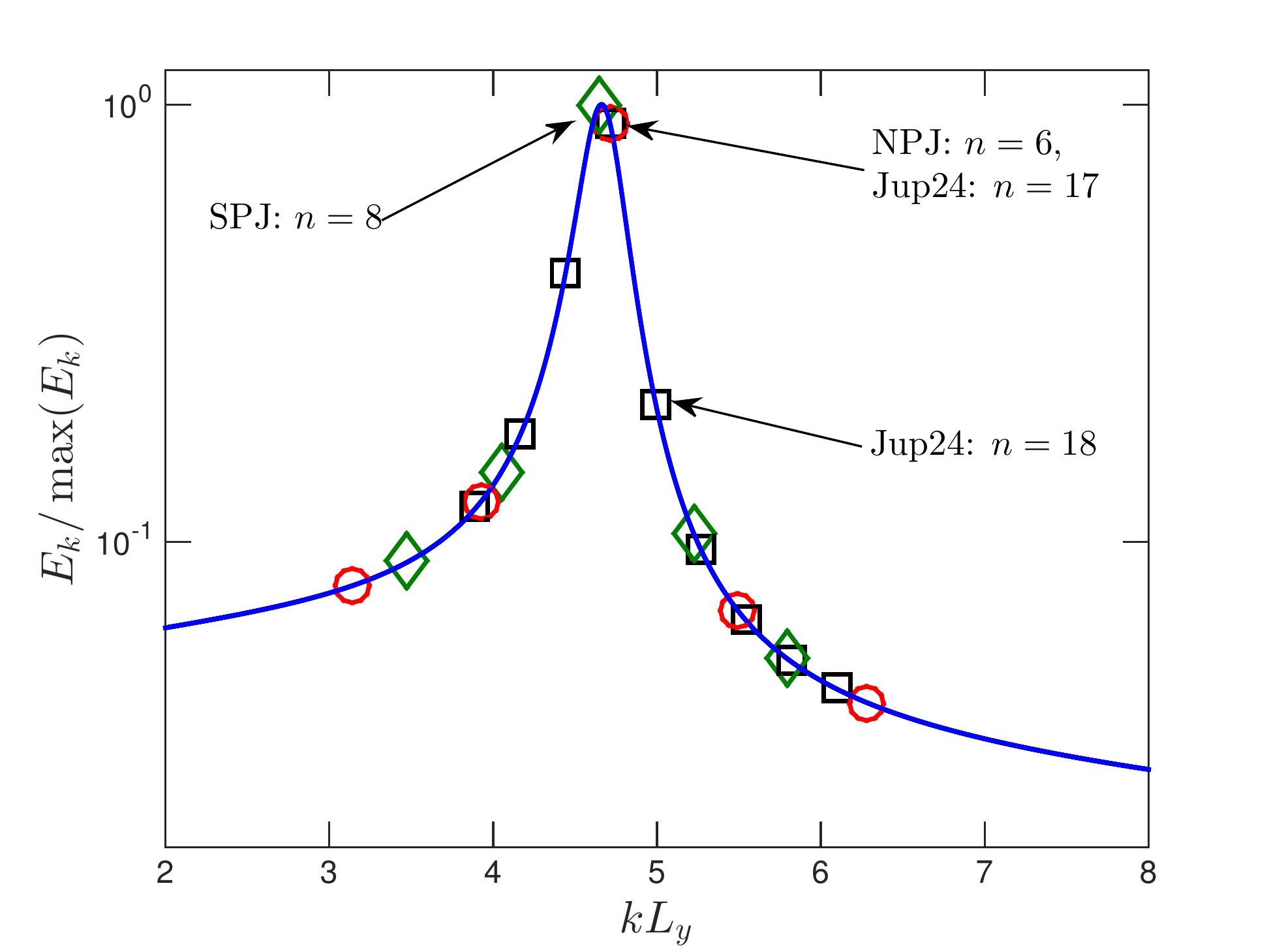}
        \caption{Normalized energy spectrum, $E_k/\max(E_k)$,
        supported by perturbations to a finite $\beta$ approximation of the universal jet
        as a function of
         zonal wavenumber $k L_y$, where $L_y$ is the width of the approximate universal
         jet. $E_k$ is the perturbation energy maintained at $k$ when the universal velocity profile, shown in Fig. \ref{fig:Unorm},
         is stochastically excited with equal energy injection rate at each wavenumber.
         The perturbation energy spectrum
        has a strong  peak at $k L_y = 4.66$.  Waves $n=4,\dots,8$ in
        Saturn's northern polar jet (with $L_y=10^3 ~\rm km$  and $k  = 2 \pi n /L_x$ with
        $L_x=80 \times 10^3 ~\rm km$) are indicated with circles. Waves $n=6,\dots,10$ in
        Saturn's southern polar jet (with $L_y=1.2 \times 10^3 ~\rm km$  and $k = 2 \pi n /L_x$ with
        $L_x=130 \times 10^3 ~\rm km$) are indicated with diamonds.
       Waves $n=14,\dots,22$ in
        Jupiter's $24^o$N jet (with $L_y=1.8 \times 10^3 ~\rm km$  and $k  = 2 \pi n /L_x$ with
        $L_x=407 \times 10^3 ~\rm km$) are indicated with squares.}
\label{fig:E_uni}
\end{figure}

\begin{figure}
	\centering
       \includegraphics[width = .5\textwidth]{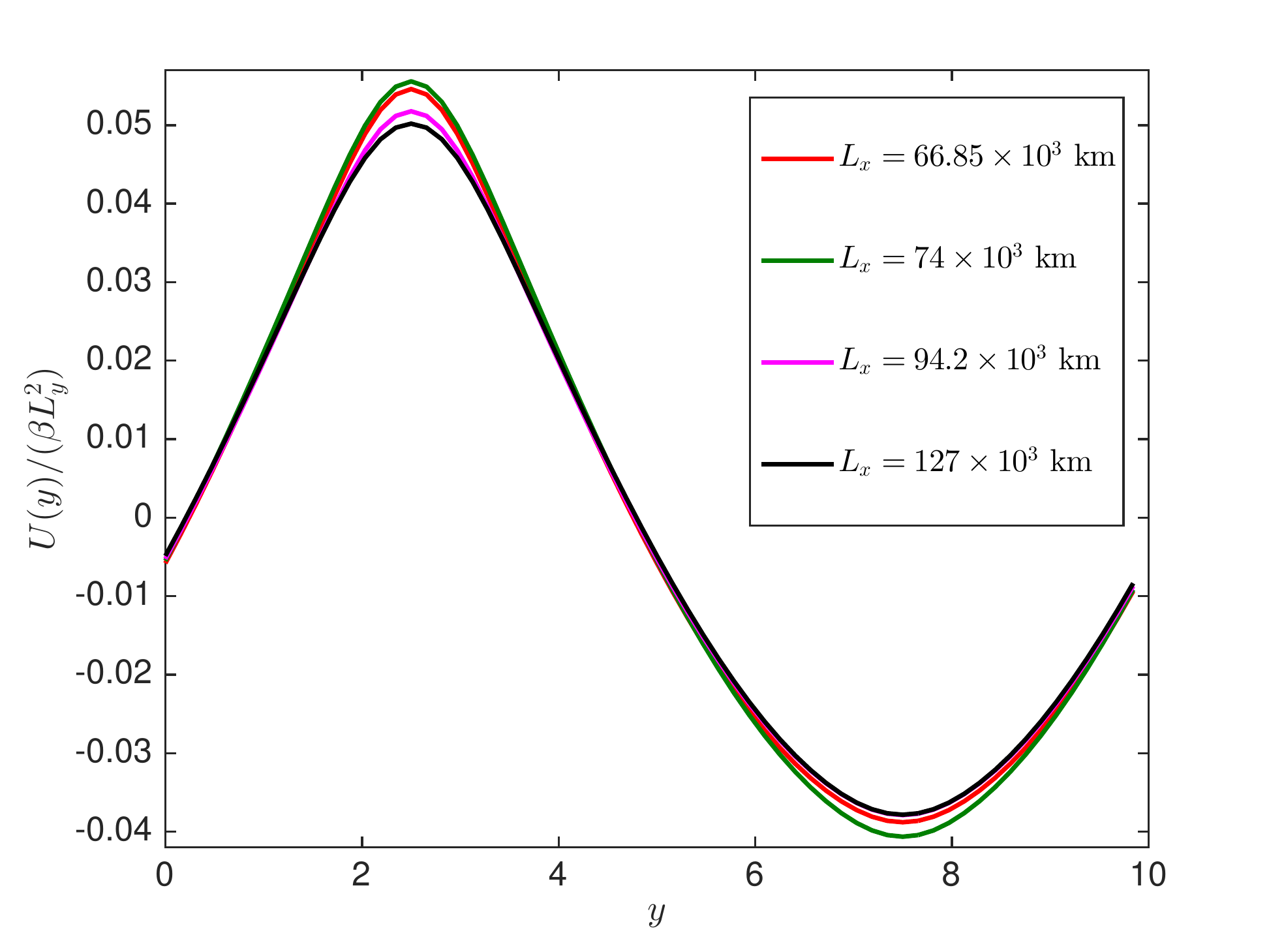}
        \caption{S3T equilibrium jets  for
        channels with $L_x = 66.85 \times 10^3~\rm km$,  $L_x = 74 \times 10^3~\rm km$,
       $L_x = 94.2 \times 10^3~\rm km$ and $L_x = 127 \times 10^3~\rm km$.  The jets are very close to the
       universal profile, despite the differences in their stability and supporting perturbation structures
       as shown in Fig. \ref{fig:kci_v} and Fig. \ref{fig:E_v}. In all cases
       $L_y=10 \times 10^3~\rm km$, $\tilde \beta= 6.9$, $r_m=0$ and $r_p = 0.2~\rm day^{-1}$.}
\label{fig:U_v}
\end{figure}
\begin{figure}
	\centering
       \includegraphics[width = .5\textwidth]{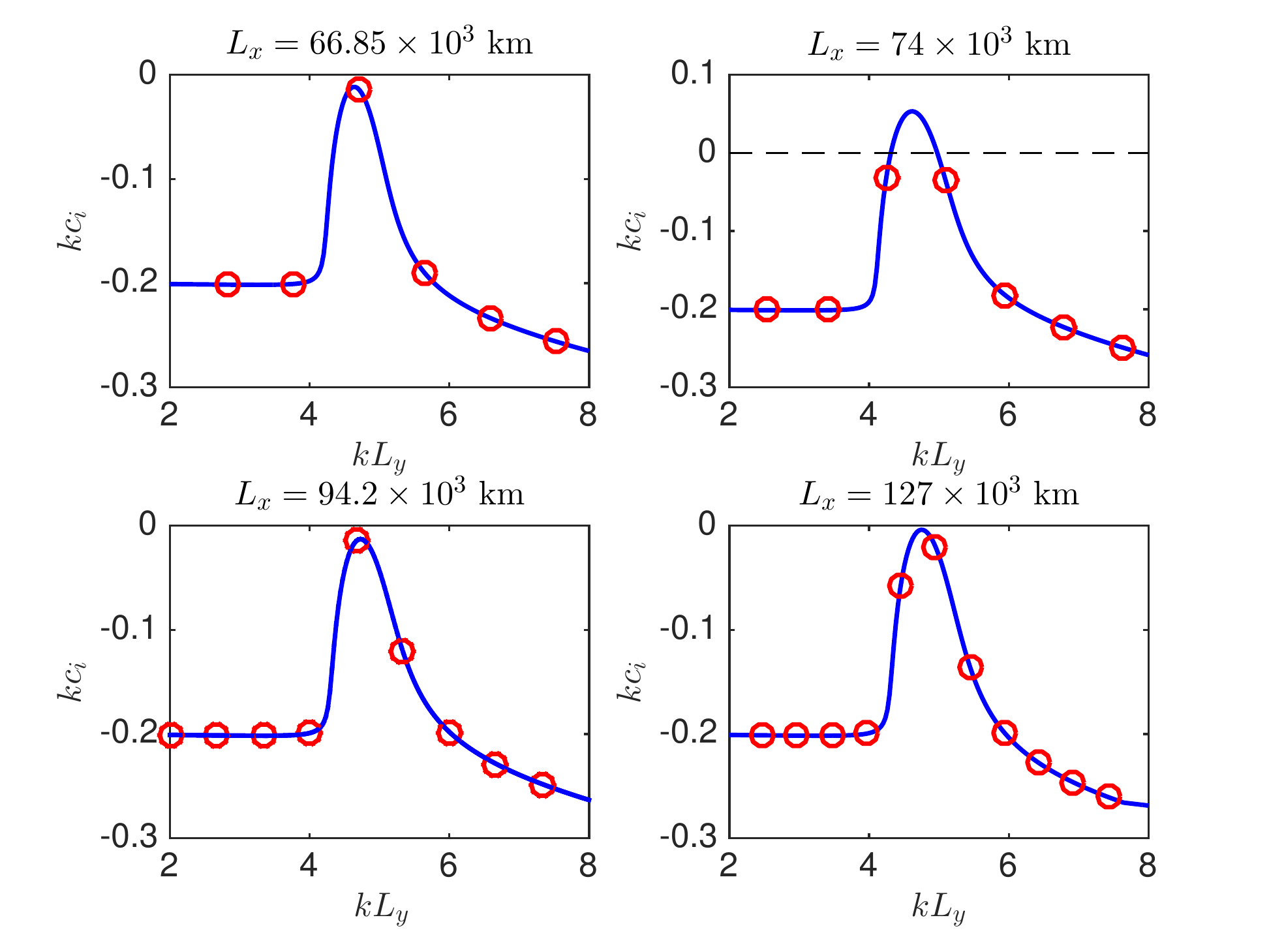}
        \caption{Maximum modal growth rate as a function zonal wavenumber $k L_y$
        for the S3T equilibrium jets shown in Fig. \ref{fig:U_v}. These jet equilibria are obtained
         for channels with $L_x = 66.85 \times 10^3~\rm km$,  $L_x = 74 \times 10^3~\rm km$,
       $L_x = 94.2 \times 10^3~\rm km$, $L_x = 127 \times 10^3~\rm km$. Circles indicate
       the growth rates
       for  the perturbations with zonal wavenumbers  that satisfy the quantization condition in each channel.
       In all cases the equilibrium jet is hydrodynamically stable at these wavenumbers, but not necessarily
       stable at all wavenumbers, as for example in the case of $L_x =74$, which has been chosen so that there is no wavenumber
       at the peak of the resonant response  where a strong instability exists.
       For $L_x= 66.85 \times 10^3~\rm km$ the jet maintains strongly
       the wave five in the channel for which $k L_y=4.695$ is close to the resonant peak.
       For $L_x= 74 \times 10^3~\rm km$ the jet maintains strongly both
       waves five and six  for which $k  L_y$ are respectively $4.24$ and $5.1$, which are removed  further than a half-width
       from the resonant peak
       (the energy at wave 6 is half that  at
wave 5). For $L_x= 94.2 \times 10^3~\rm km$ the jet maintains strongly
       wave seven in the channel for which $k L_y=4.67$ is close to the resonant peak. Finally,
       For $L_x= 127 \times 10^3~\rm km$ the jet maintains strongly
       waves nine and ten (with  energy at wave 10  slightly higher
       than the energy of wave 9) for which $k  L_y$ are respectively $4.45$ and $4.94$.  Parameters as in Fig. \ref{fig:U_v}.}
\label{fig:kci_v}
\end{figure}

\begin{figure}
	\centering
       \includegraphics[width = .5\textwidth]{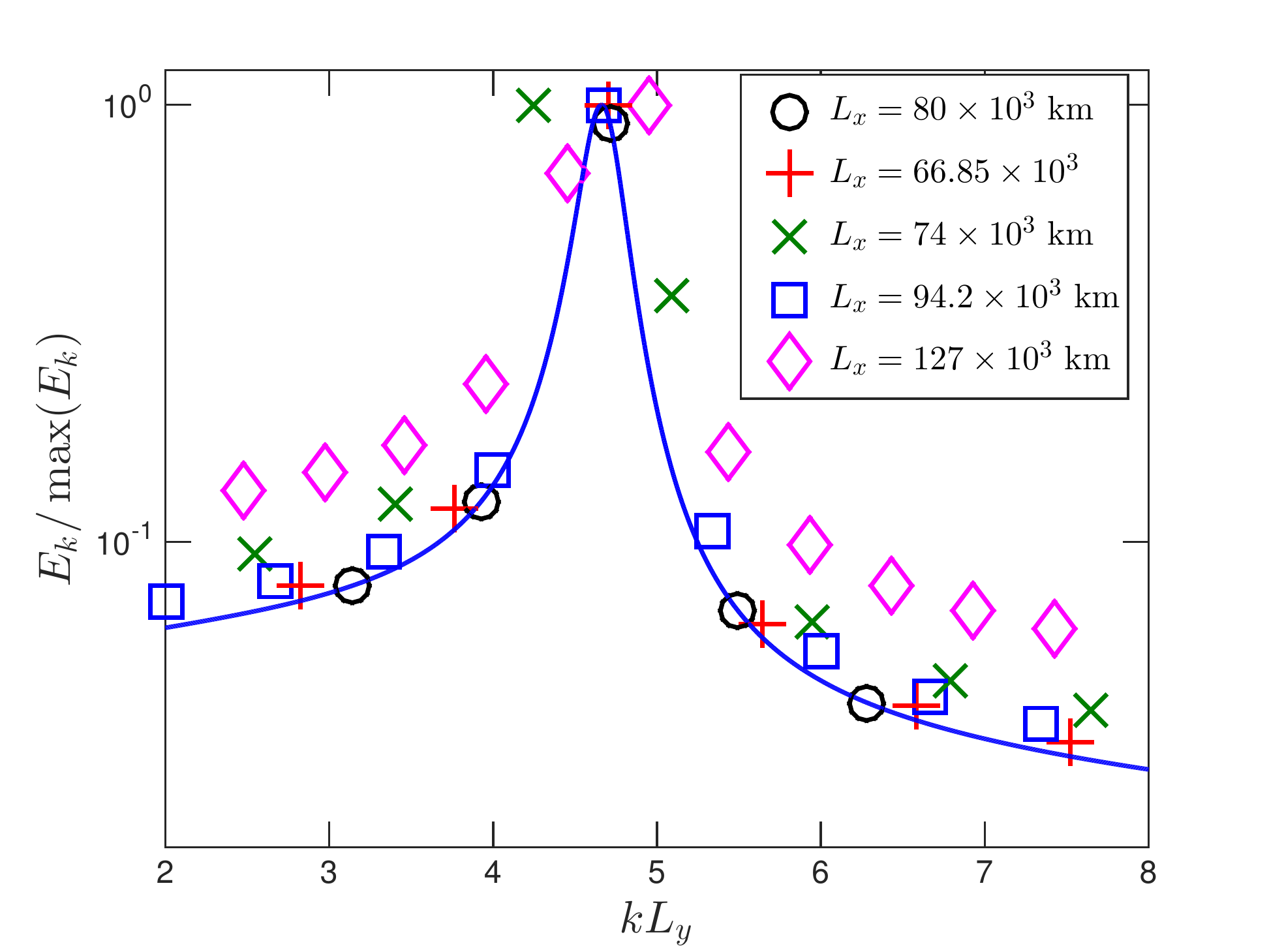}
        \caption{Comparison of  the normalized energy spectrum, $E_k/\max(E_k)$  supported by the  equilibrium jets of
        Fig. \ref{fig:U_v}  to the  energy spectrum supported by the approximation to the universal profile. Parameters as in Fig. \ref{fig:U_v}.}
\label{fig:E_v}
\end{figure}

\begin{figure}
	\centering
       \includegraphics[width = .5\textwidth]{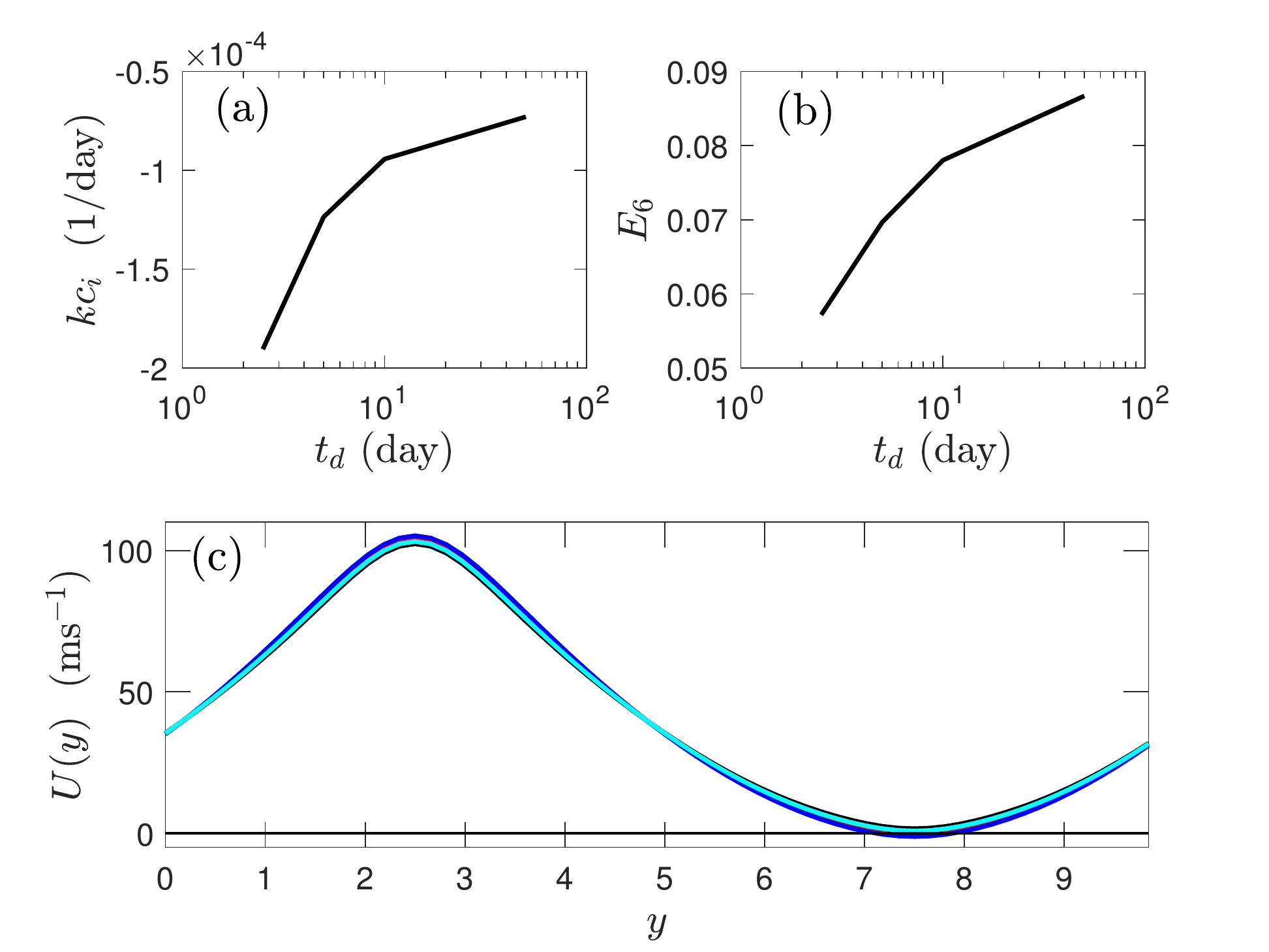}
        \caption{Dependence of the S3T equilibrium jet and associated wave six
        amplitude and decay rate on the dissipation rate $r_p$  of the small scale waves. Shown in (a) is the decay rate of the least
        damped mode 6 as a function the dissipation time $t_d = 1/r_p \rm (day)$ imposed on
          waves $n>10$. Waves $n \le 10$ are dissipated with $t_d = 5 ~\rm day $. This figure shows that as the dissipation
          rate of the small scale waves with $n>10$,  which are responsible for producing the upgradient fluxes forcing the jet,
          decreases,  and the energy at these scales consistently increases,  wave six approaches neutrality
          while its energy, shown in (b), increases. The corresponding equilibrium barotropic jet for
           $t_d= 2.5, 5, 10, 50~\rm day$ of the $n>10$ waves is shown in (c). This figure shows the insensitivity of the equilibrium jets to the
           small scale wave dissipation rate. In these simulations the energy input to
            each of the waves $n \le 10 $ is  $1 \%$ of the energy input to each
            of the waves with $n > 10$.  Other parameters $L_x = 80 \times 10^3~\rm km$,
       $L_y=10 \times 10^3~\rm km$, $\tilde \beta= 6.9$, $r_m=0$.}
\label{fig:E_disip}
\end{figure}

\begin{figure}
	\centering
       \includegraphics[width = .5\textwidth]{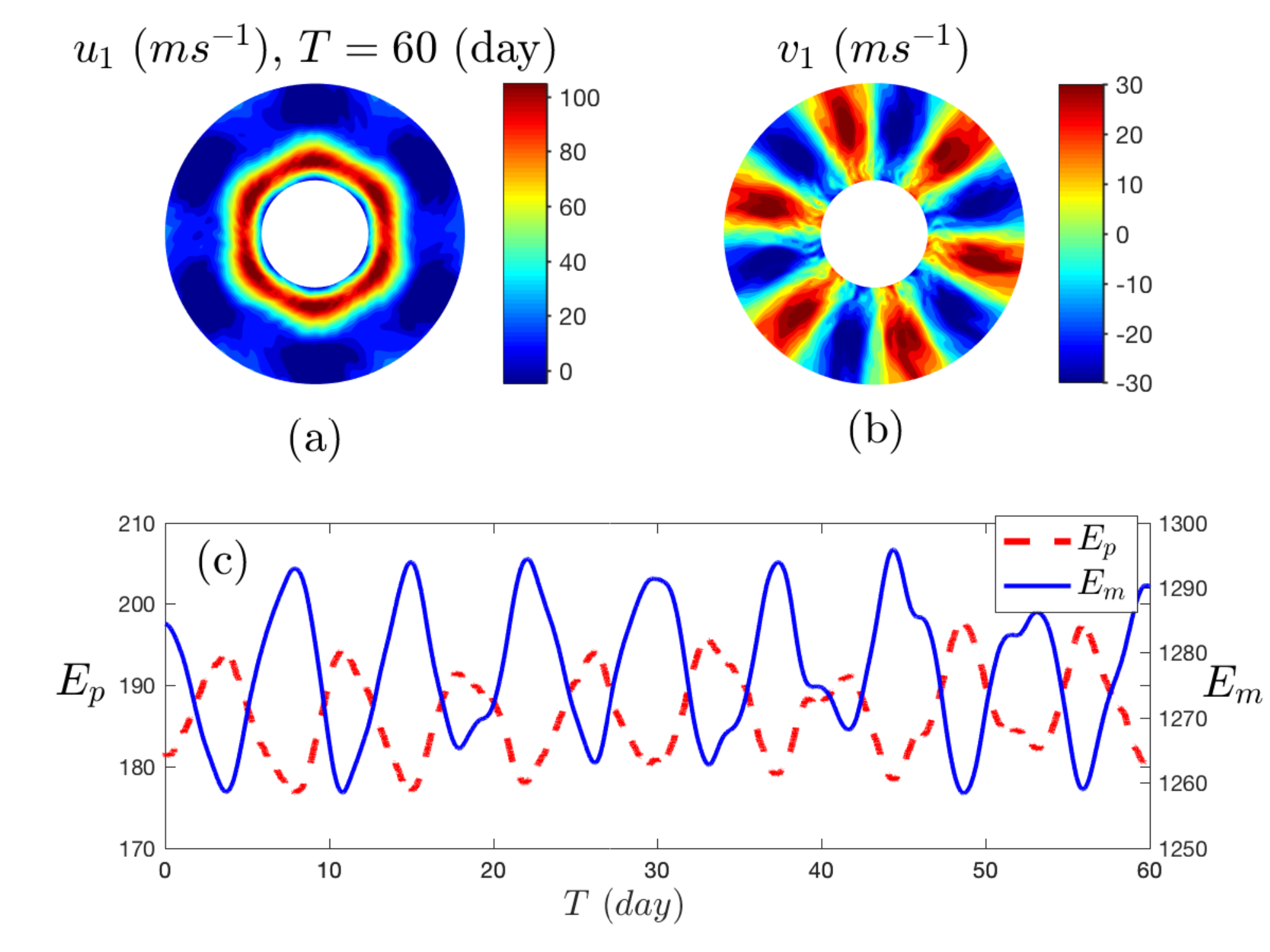}
        \caption{Snapshot  of
        the flow  in the upper layer obtained
         from a stochastic simulation of the QL equations  with
         parameters that reproduce the observations. In this polar plot the external circle
         of the annulus  corresponds to latitude $70^o$ and the inner circle to  $82^o$.
         Shown are:  contours of
         the total top layer
         zonal velocity, $u_1$, (a),
          the meridional velocity, $v_1$, (b),  and a time series of the kinetic energy of the non-zonal
component, $E_p$, and of the mean
flow, $E_m$. The flow is close to barotropic. The temporal evolution of this
simulation can be seen in
          the movie  included in  the
Supplemental Material.    The vacillation  in  the mean and
perturbation energy, the perturbation component of which
is predominantly concentrated at  wave 6,
reveals a compensating  exchange of energy between wave 6 and the mean flow, which can be identified with
the least damped S3T mode of the jet equilibrium
excited by the fluctuations of the QL simulation (see Fig. \ref{fig:S3T_conf}).
In these simulations the large scales waves ($n\le 10$) are dissipated at rate $r_p=0.005~\rm day^{-1}$
 while small scales waves ($n > 10$) are dissipated at rate $r_p=0.05~\rm day^{-1}$. The total energy input
 is $1~\rm W m^{-2}$  distributed equally among waves $n>10$
            (the energy input to
            each of the waves $n \le 10 $ is  $ 10^{-4}$ of the energy input to each
            of the waves with $n > 10$).    Other parameters $L_x = 80 \times 10^3~\rm km$,
       $L_y=10 \times 10^3~\rm km$, $\tilde \beta= 8.6$, $r_m=0$. }
\label{fig:E_vs}
\end{figure}

\begin{figure}
	\centering
       \includegraphics[width = .5\textwidth]{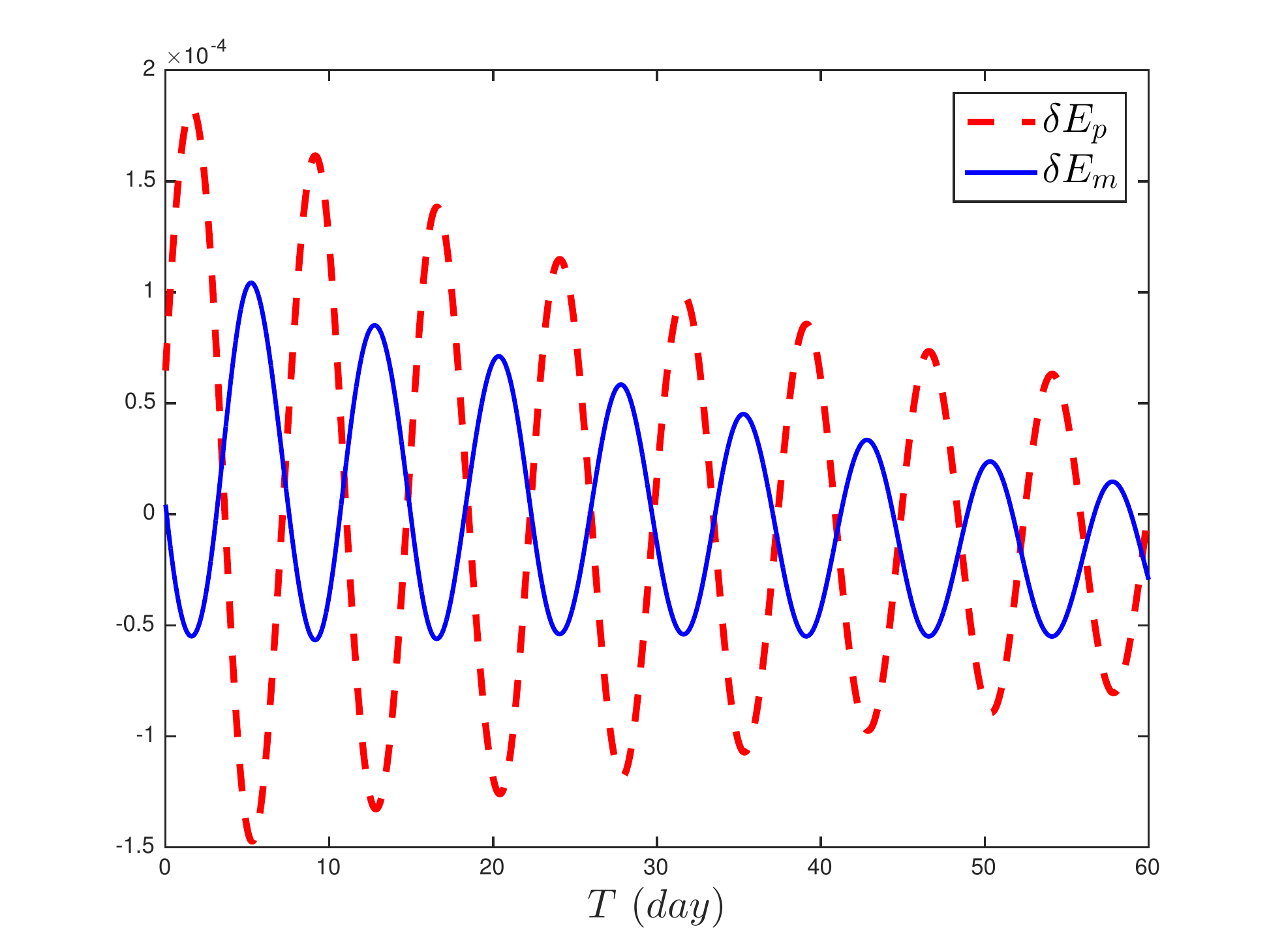}
        \caption{Oscillations resulting from perturbation to the S3T zonal jet equilibrium
        for the parameters in Fig. \ref{fig:E_vs}. Shown  is the time evolution of  the
         perturbation energy fluctuation from the equilibrium value, $\delta E_p$, (solid)  and fluctuation of
         the mean energy  as a departure from its equilibrium value, $\delta E_m$. The equilibrium covariance of  the
        S3T jet equilibrium was perturbed with a  random perturbation  sufficiently small to render the
         dynamics of relaxation to the S3T equilibrium linear. The
        decaying oscillations in the energy are shown after an initial adjustment has occurred.
        The frequency of the oscillations matches the oscillations shown in the QL simulation in Fig. \ref{fig:E_vs} and the oscillations in this figure are interpreted as revealing the least
        damped S3T mode about the S3T equilibrium which are being stochastically
        maintained  by the fluctuations  in the
        QL simulation.  }
\label{fig:S3T_conf}
\end{figure}

\section{Discussion}

Statistical state dynamics at second order
predicts that a stochastically excited  two  layer fluid
in a meridionally confined channel with vanishing
meridional temperature gradient and jet scale dissipation supports barotropic jets
that asymptotically  approach a universal structure as the channel length $L_x \rightarrow \infty$ and
$\beta\rightarrow \infty $.
The velocity of this universal jet
scales  as  $\beta L_y^2$.  Saturn's  $74^o$N jet, Saturn's  $70^o$ S  jet and
Jupiter's $24^o$N  jet    closely approximate this universal structure.
Associated with this barotropic mean jet
is  a  universal equilibrium
covariance   satisfying   the non-dimensional  barotropic component of  \eqref{eq:pertCe}:
\begin{equation}
 {\tilde \A}_{\tilde k}^{\psi \psi}(\tilde{U}) \, \C_{\tilde k}^{\psi \psi} +
 \C_{\tilde k}^{\psi \psi} \,{\tilde \A}_{\tilde k}^{\psi \psi \dagger}(\tilde{U}) = - \tilde{\varepsilon} \Q_{\tilde k}^{\psi \psi}~,\label{eq:bar1}
\end{equation}
with
\begin{equation}
\tilde{\A}_{\tilde k}^{\psi \psi} =\tilde{\Delta}_{\tilde k}^{-1} \left [ -\ii \tilde{k}  \tilde{U}
 \tilde{\Delta}_{\tilde k} -\ii {\tilde k} \left (
1 - \tilde{D}^2 \tilde{U} \right ) \right ]  -\tilde{r}_p~,
\end{equation}
in which length is scaled by $L_y$,  time by  $1/(\beta L_y)$ and
velocity  by $\beta L_y^2$,
so that
${\tilde k} = k L_y$, $\tilde{D}^2 = L_y^2 D^2$,
$\tilde{\Delta}_{\tilde k}= \tilde{D}^2 -{\tilde k}^2$,
$\tilde{U} = \beta L_y^2$,  $\tilde{r}_p = r_p / (\beta L_y)$ and $\varepsilon = \varepsilon / (\beta L_y)$.
In the limit $\beta \rightarrow \infty$
both  $\tilde{r}_p$ and $ \tilde{\varepsilon} $ can vanish while their ratio, $\tilde{\varepsilon}/\tilde{r}_p $,
is finite.
A finite $\beta$ approximation to this
universal energy spectrum  of the perturbation field  associated with this universal jet structure
is  shown in Fig. \ref{fig:E_uni}. It has
a single  peak at $\tilde{k}= 4.66$ that arises in association with
a neutral wave with phase velocity inside the retrograde jet.
The universality of the profile and of the resonant response is
predicated on assuming a zonally unbounded channel, $L_x \rightarrow \infty$.
Channels with finite  zonal extent
make available to the dynamics only a discrete set of  zonal wavenumbers.
If quantization conditions allow a wave or waves with    $\tilde{k}  =  2 \pi n L_y / L_x \approx 4.66$,
where $n$ is a discrete wavenumber in the periodic channel of length $L_x$,
then  the wave(s) with $n$
nearest to $4.66 L_x / (2 \pi L_y)$ will equilibrate at highest amplitude.
Given that jets with near universal structure are  observed at different latitudes
in the outer planets (e.g. the northern and southern polar jets on Saturn and the $24^o$N jet on Jupiter)
we are led to predict   that observation of the perturbation field at these locations will
reveal the resonant Fourier components associated with  the predicted
resonant wave(s),  although  these waves may be  incoherent or weak if the jet forcing is weak. Predictions of
the resonant wavenumbers for Saturns NPJ, SPJ and Jupiter's $24^o$N jet are indicated in Fig. \ref{fig:E_uni}.
Moreover, in the case of  jets that are  located near the pole, the decrease in the  length of the latitudinal circle,  $L_x$,
results in substantial separation of the allowed modes on the resonant curve which favors  prominent appearance
of the mode  nearest to resonance.

We next investigate the impact of quantization of the zonal wavenumbers
associated with variation in channel length, $L_x$, on the S3T equilibrium
jet and the excitation of the resonant or near resonant waves.
We choose to compare the equilibria in channels
with $L_x = 66.85  \times 10^3~\rm km$ and $L_x = 94.2  \times 10^3~\rm km$
in which  waves 5 and 7 are exactly resonant with   the equilibria in
channels $L_x=74 \times 10^3~\rm km$  and  $L_x=127 \times 10^3~\rm km$
in which two waves are near resonant. The equilibrium jet is found to be
little modified (see Fig. \ref{fig:U_v}).  In all cases  the equilibrated profile is hydrodynamically stable for
perturbations at the allowed  wavenumbers  (see Fig. \ref{fig:kci_v})  and the resonant or near resonant waves
are  strongly maintained as shown in
Fig. \ref{fig:E_v}.

S3T integrations have demonstrated that the  equilibium jets are only
weakly dependent  on energy input rate, on the vertical structure of the forcing
and on the dissipation time scale  of the small scale structures.
We show further here that equilibrium jets with approximate universal structure
are also insensitive to  variations in the amplitude of the excitation of the  resonant
waves associated with jet equilibration.
The amplitude of the resonant wave(s) is regulated so that at equilibrium
they dissipate the  energy transferred to the jet by the smaller scales.
Consequently,   as  shown in Fig. \ref{fig:E_disip}, as the dissipation, $r_p$,  of the small scales
is decreased  while holding the energy input rate constant, the large scale resonant waves
responsible for regulating the jet to equilibrium approach neutrality so that
their amplitude increases sufficiently to produce jet damping equal
to the energy input rate from the small scale waves, while the mean jet amplitude remains essentially  unaffected.
In the  calculations shown
in Fig. \ref{fig:E_disip} we have chosen to excite each of the waves $n=1,\cdots, 12$ with energy input  $1\%$ of the energy input in each of the waves
$n>12$ in order to demonstrate that the same equilibrium jet profile is  obtained
if the large waves were negligibly excited.  This is because these waves obtain
their energy primarily through non-normal interaction with the jet.

The amplitude of the primary resonant wave is that required to dissipate the energy coming into
the jet from  shearing by the jet of the small scale waves.
This amplitude is proportional to the  small scale wave excitation and inversely proportional to the dissipation
rate of the resonant wave.  Observations of the NPJ reveal this amplitude and the SSD theory allows us to predict
parameter values compatible with these observations.
However, observations of the wave also reveal its coherence which further constrains
parameter values.  The SSD theory developed above exploits the assumption of an
infinite ensemble of perturbations so that the fluctuations
of the perturbations have been suppressed by the Central Limit Theorem.
This simplification is necessary for the theoretical development but
by relaxing this infinite ensemble assumption we can obtain predictions for the variance of wave
six fluctuations as a function of system parameters
and thereby further constrain these parameters.   While we do not have data on the temporal variations
of the flow fields at wave six, we do know from photographs of the NPJ that wave six is  coherent
and slowly varying.  We can constrain the values for the energy input
at the smaller scales and their dissipation by going beyond the structure of the flow at statistical equilibrium
to actual simulations of the stochastically forced QL equations
 \eqref{eq:QLbrcl} coupled with the mean equations   \eqref{eq:NL_mbrcl}
 in order to determine parameters consistent with the observed coherence of the hexagonal pattern.
 Although other parameter choices
 could be made we  obtained  agreement with observations of the steadiness of wave six using
 the simple two layer model with $\tilde \beta = 8.6$.
A snapshot of the fields obtained from such a simulation is shown in  Fig. \ref{fig:E_vs}
            and a movie of the evolution of the upper level fields in this simulation can be seen in the supplemental materials.
            The simulation spans 60 days and the movie is seen in a frame of reference which rotates with the maximum retrograde
            speed of the jet at S3T equilibrium. The wave 6  phase speed  is almost equal to this speed and it can be seen in
            the movie to be both stationary and also coherent.  There are small fluctuations, caused by the random excitation in the
            QL simulation, in both the perturbation energy (mainly wave 6) and the energy of the mean flow. These oscillations are
            compensating and it is interesting to note that they can be given an analytical interpretation: they are the
        least damped S3T mode about the S3T equilibrium, which is excited by the fluctuations  in the
        QL simulation. Indeed,  this decaying mode can be identified by using an S3T simulation, which is free of fluctuations,
        in which the S3T equilibrium is perturbed and the approach to equilibrium is plotted,  as
        shown in Fig. \ref{fig:S3T_conf}. A similar identification of the latent jet fluctuations seen in the oceans as
        the S3T damped modes  of the   uniform equilibrium  excited by the fluctuations was proposed in Constantinou et al.~\citep{Constantinou-etal-2014}.

\section{Conclusion}

Large-scale coherent structures such as jets, meandering jets
are characteristic features of turbulence in planetary
atmospheres. While conservation of energy and enstrophy in inviscid 2D
turbulence predicts spectral evolution leading to concentration of
energy at large scales, these considerations cannot predict the phase
of the spectral components and therefore can neither address the central
question of the organization of the energy into
 jets with specific structure nor the existence of the coherent component of
the planetary scale waves. In order to study structure formation
additional aspects of the turbulence dynamics beyond conservation principles
must be incorporated in the analysis.  SSD models have been developed
to study turbulence dynamics and specifically to solve for turbulent state equilibria
consisting of coexisting coherent mean structures and incoherent turbulent components
which together constitute the complete state of the turbulence at second order.
In this work a second order SSD of a two-layer baroclinic model was used to study the
jet-wave-turbulence coexistence regime in Saturn's NPJ.  This second order SSD model, referred to as the S3T model,
is closed by a stochastic  parameterization that accounts for
both the neglected nonlinear dynamics of the perturbations from the zonal mean
as well as the excitation maintaining the turbulence. The equation for the zonal mean
retains its interaction through Reynolds stress with the perturbations.

In this model a jet forms as an instability  and grows at first exponentially eventually equilibrating
at finite amplitude.  Exploiting the simplicity of the asymptotic regime in which the jet is undamped
makes it possible to obtain a universal jet structure and jet amplitude scaling.
Given that the associated jet structure and its amplitude scaling is robust in the SSD model
we  conclude that  the observed structure of the NPJ can only be maintained as an equilibrium
state with  a value of $\beta$ greater than the planetary value.
This requirement implies existence of a topographic beta effect with a specific predicted value.  Incorporating
the implied poleward decreasing stable layer depth into the model results in the model producing
the observed jet structure.
In the model a stable retrograde mode  of the Rossby wave spectrum with wavenumber six
becomes neutrally stable as the jet amplitude increases under Reynolds stress
forcing by the small scale turbulence
increased and by inducing strong non-normal
interaction with the jet this wave six arrests its growth via perturbation Reynolds stresses.
This composite structure equilibrates in the form of a hexagonal  jet in agreement with the NPJ observations.
Among the  correlates of this theory is the predicted existence of the observed
robust vorticity gradient reversals
in both the prograde and retrograde jets as well as the location and structure of these reversals.



\begin{acknowledgments}
Brian Farrell was partially supported by NSF AGS-1246929. We thank Navid Constantinou for helpful discussions.
We also thank the anonymous reviewer for his comments that have improved the paper.
\end{acknowledgments}


%

\end{document}